\documentclass[a4paper,twocolumn]{article}
\usepackage{cite}

\usepackage[pdftex]{graphicx}
\usepackage[font=footnotesize]{caption}
\usepackage{amsmath}
\usepackage{amsfonts}
\usepackage{amsthm}
\usepackage{dsfont}

\usepackage{array}

\newcommand{\CC}{\mathds{C}}
\newcommand{\RR}{\mathds{R}}
\newcommand{\ud}{\mathrm{d}}

\newcommand{\rmi}{{\rm{i}}}
\newcommand{\rme}{{\rm{e}}}

\renewcommand{\Im}{\operatorname{Im}}
\renewcommand{\Re}{\operatorname{Re}}

\begin{document}
\title{Propagation of Gaussian beams in the presence of gain and loss}
\author{Eva-Maria~Graefe, Alexander Rush, and Roman~Schubert
\thanks{Eva-Maria Graefe and Alexander Rush are with the Department
of Mathematics, Imperial College London, London, SW7 2AZ, United Kingdom}
\thanks{Roman Schubert is with the School of Mathematics, University of Bristol, Bristol BS8 1TW, United Kingdom}}

\maketitle
\begin{abstract}
We consider the propagation of Gaussian beams in a waveguide with gain and loss in the paraxial approximation governed by the Schr\"odinger equation. We derive equations of motion for the beam in the semiclassical limit that are valid when the waveguide profile is locally well approximated by quadratic functions. For Hermitian systems, without any loss or gain, these dynamics are given by Hamilton's equations for the center of the beam and its conjugate momentum. Adding gain and/or loss to the waveguide introduces a non-Hermitian component, causing the width of the Gaussian beam to play an important role in its propagation. Here we show how the width affects the motion of the beam and how this may be used to filter Gaussian beams located at the same initial position based on their width.
\end{abstract}

\section{Introduction}
While quantum dynamics is in general very different from classical dynamics, for short times initially Gaussian wave packets stay approximately Gaussian and follow classical trajectories. This has been shown by Hepp and Heller using the fact that Gaussian states stay exactly Gaussian for harmonic Hamiltonians, and Taylor expanding more general Hamiltonians up to second order around the centre of the wave packet \cite{HeppHeller} (see also \cite{Litt86} for a review on the properties and dynamics of Gaussian wave packets) . As long as the wave packet stays well localized compared to the scale on which the higher order terms in the Taylor expansion of the Hamiltonian are negligible this is a good approximation. It is also the main ingredient of more powerful methods to approximate dynamics in the deep quantum regime such as the frozen Gaussian propagator and the method of hybrid mechanics \cite{Hell_dyn}.

There are many analogies between the wave nature of light and the quantum mechanical wave function. In particular, the propagation of light in wave guides in the paraxial approximation is described by a time-dependent Schr\"odinger equation (where the propagation direction acts as time). This has led to spectacular demonstrations of quantum dynamical effects, as well as to new applications  in optics (see, e.g. \cite{Long09b} and references therein). In recent years, there has been a surge of interest in optical waveguides with gain and loss, mimicking non-Hermitian, and in particular $PT$-symmetric \cite{PT}, quantum systems \cite{PT_optics}. It is an interesting question what might be the fate of initial Gaussian wave packets in the limit of large locally approximately harmonic potentials in the spirit of Hepp and Heller. In \cite{11nhcs} this question has been investigated, and a new type of classical dynamics guiding the centers of Gaussian wave packets has been derived. In this dynamics the changing shape of the wave packet couples back into the dynamics of the centre. That is, even in the limit where the beam width is completely negligible compared to the external potential and where the dynamics is accurately described by the Gaussian approximation, there are many classical trajectories for a given set of initial position and momentum, depending on the width of the wave packet. 

Here we review the approximation and give explicit equations of motion for the special case of a Hamiltonian with complex potential, and investigate the resulting dynamics in an example of a $PT$-symmetric multimode waveguide. The approximation accurately describes the propagation and effects such as the typical power oscillations in PT-symmetric systems if the potential is chosen large enough compared to the beam width associated to the ground state. We demonstrate how the shape of the beam affects the propagation, and how this might be used as a filtering device.

\section{Gaussian beams in the paraxial wave equation}
For simplicity we consider the propagation of light in a two-dimensional waveguide where the (complex) refractive index $n$ is constant along the $z$-direction and varies along the $x$-direction. The results are straight forwardly generalised to higher dimensions and refractive indices that are modulated in $z$-direction as well. 

In the paraxial approximation the propagation of the electric field amplitude $\psi$ is described by the Schr\"odinger equation 
\begin{equation}
\rmi\hbar\frac{\partial\psi}{\partial z}=-\frac{\hbar^2}{2n_0}\frac{\partial^2\psi}{\partial x^2}+V(x)\psi,
\end{equation}
where $\hbar:=\frac{\lambda}{2\pi}$, and $\lambda$ is the wavelength, and where the effective potential
\begin{equation}
V(x)=\frac{n_0^2-n^2(x)}{2n_0}\approx n_0-n(x)
\end{equation}
is approximately given by the refractive index profile $n(x)$ of the waveguide, where $n_0$ is the reference refractive index of the substrate. Without loss of generality we use rescaled units with $n_0=1=\hbar$ in what follows. 

We consider the propagation of an initial Gaussian beam of the form
\begin{equation}
\psi(x,0)=\left(\tfrac{{\rm Im}\left(B_0\right)}{\pi}\right)^{\frac{1}{4}}{\rm e}^{ \rmi\left(\frac{B_0}{2}\left(x-q_0\right)^{2}+p_0\left(x-q_0\right)\right)},
\label{eq:initial}
\end{equation}
where $q_0\in\RR$ is the position of the center of the beam, $p_0$, the expectation value of the quantum momentum operator, describes the angle of incidence, and $B_0\in\CC$ encodes the shape of the beam. In particular it holds $\Delta q_0=\frac{1}{\sqrt{2\Im B_0}}$, and $\Delta p_0=\frac{|B_0|}{\sqrt{2\Im B_0}}$, and we demand $\Im(B_0)>0$ such that the wave packet is normalisable. 

In the spirit of Hepp and Heller we make a Gaussian ansatz for the propagated beam
\begin{equation}
\psi(x,z)\!=\!N\!(z)\!\!\left(\!\tfrac{{\rm Im}\left(B(z)\right)}{\pi}\!\right)^{\frac{1}{4}}\!\!\rme^{\rmi\left(\!\frac{B(z)}{2}\left(x-q(z)\right)^{2}+p(z)\left(x-q(z)\right)+\alpha(z)\!\right)},
\label{eq:C_S_wavepacket}
\end{equation}
with $q(z),p(z)\in \RR$, $B(z)\in\CC$. Here $N(z)\in\RR$ denotes the norm, and $\alpha(z)\in\RR$ is an additional phase. We Taylor expand the potential around the center $q(z)$ of the wave packet and equate terms of the same orders
of $\left(x-q\right)$ to find the dynamical equations for the parameters \cite{11nhcs}
\begin{equation}
\begin{aligned}
\dot{p} = &-V_R'(q) + \frac{\Re(B)}{\Im(B)} V'_I(q), \\
\dot{q} = &p + \frac{1}{\Im(B)} V'_I(q),\\
\dot B=& - B^{2} - V^{\prime\prime}_R(q) - \rmi V^{\prime\prime}_I(q),\label{eq:Gaussian_approx}\\
\dot{N} =& \left(\frac{1}{\hbar}V_I(q) + \frac{1}{4\Im(B)}V''_I(q)\right)N,\\
\dot \alpha=& p\, \dot q-\frac{p^2}{2}-V_R(q)-\frac{1}{2}\Im(B).
\end{aligned}
\end{equation}
Here $V_R$ and $V_I$ denote the real and imaginary part of the (negative) refractive index, respectively. The dot denotes a derivative with respect to $z$, and the prime a derivative with respect to $q$. To keep the notation compact we have dropped the explicit dependence on $z$ of the parameters, and we shall continue to do so in what follows.

\begin{figure}[t]
\centering
\includegraphics[width=0.24\textwidth]{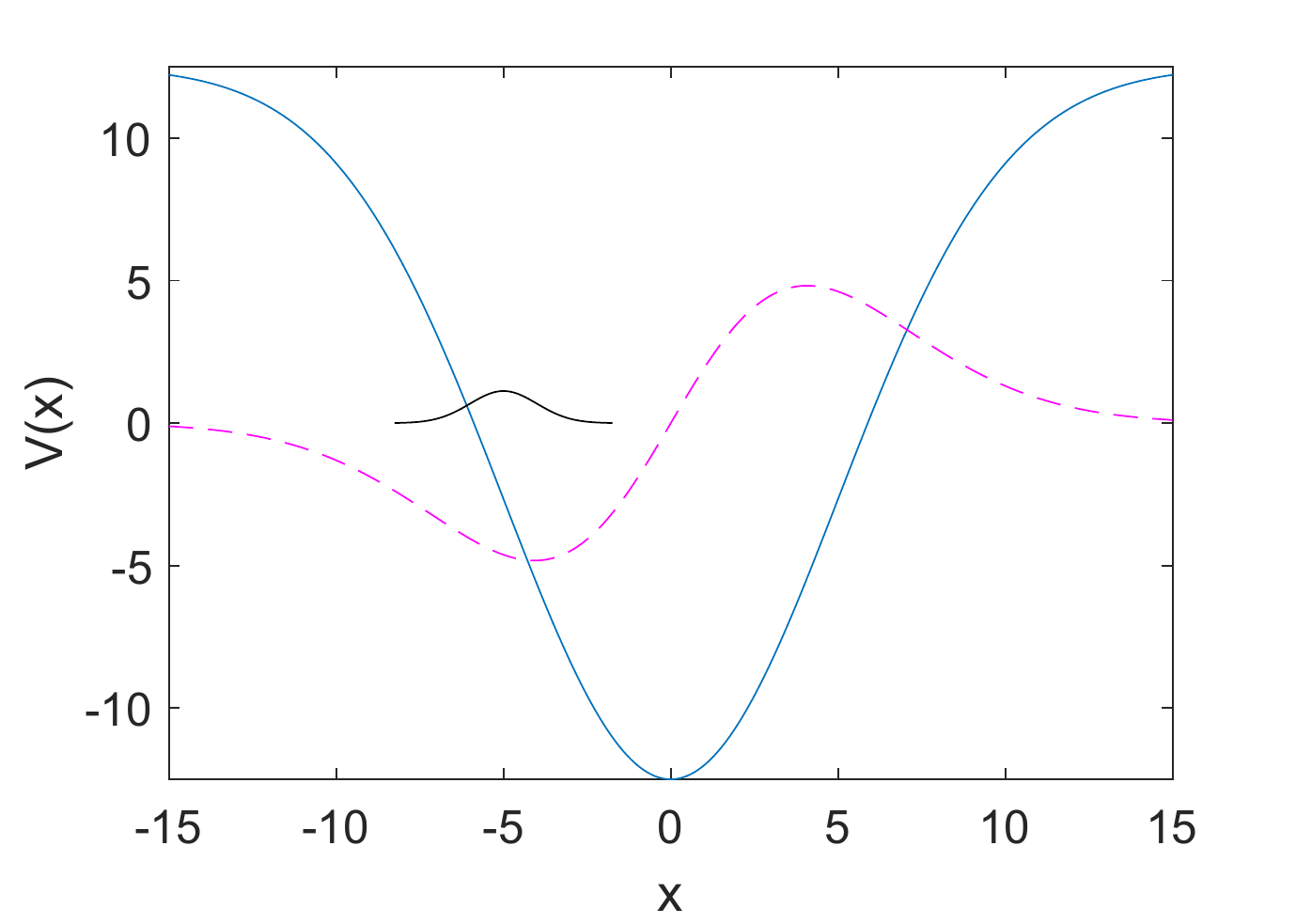}
\includegraphics[width=0.24\textwidth]{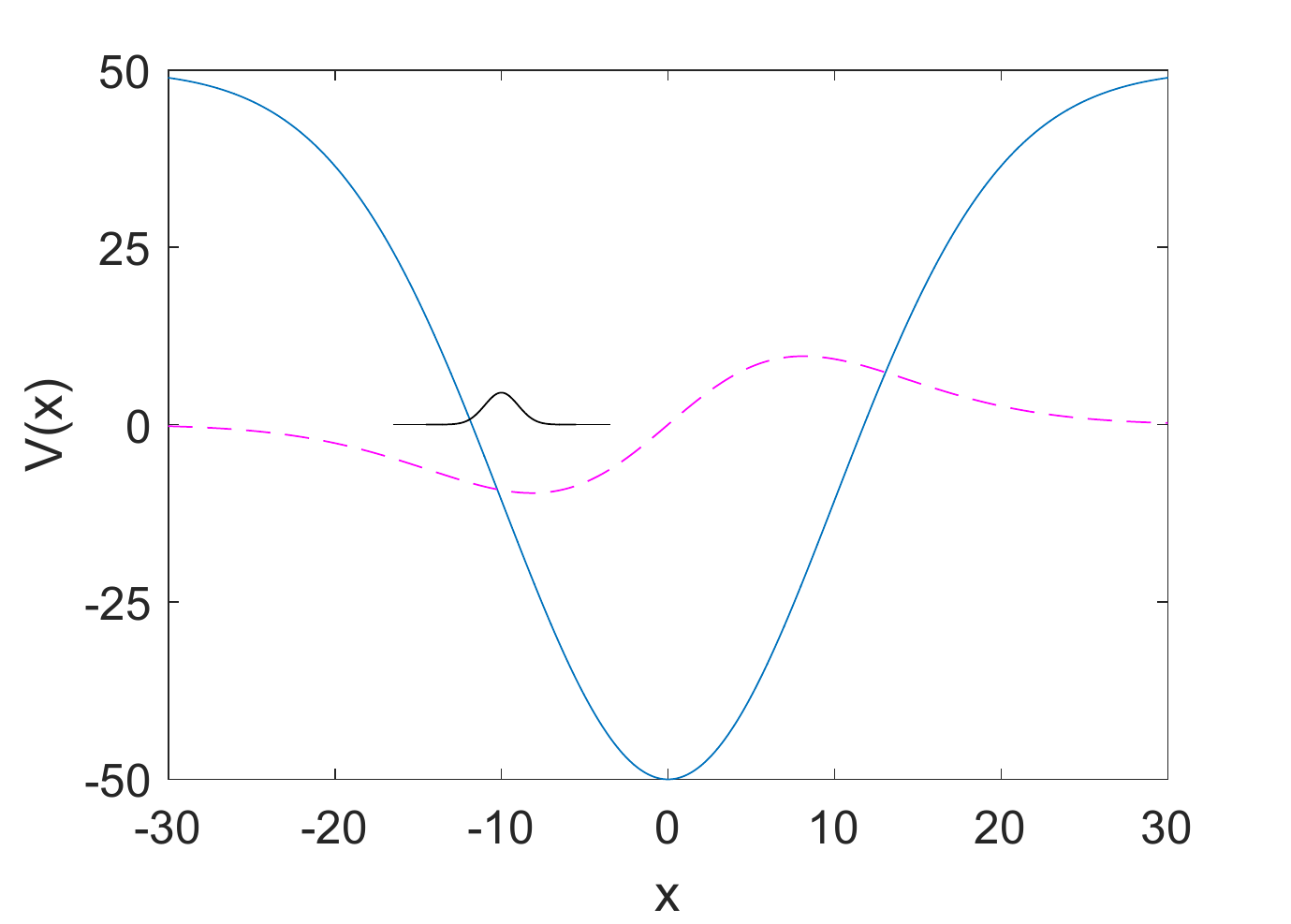}
\caption{Real (solid blue lines) and imaginary (dashed magenta lines) parts of the potential (\ref{eq:pot_ex1}) with $\gamma=1$, $\omega=1$, and $\eta=5$ (on the left) and $\eta=10$ (on the right). The imaginary part is scaled up by a factor of two, to make it more visible. For comparison we also show the probability distribution of a Gaussian beam with $B=\rmi$, corresponding to the width of the approximate ground state of the real part of the potential.}
\label{fig_pot1}
\end{figure}

The first two equations in (\ref{eq:Gaussian_approx}) are the generalisations of the standard Hamilton's equations for the dynamics of the position and the momentum to systems with gain and loss. In stark contrast to the well-known conservative case these equations depend explicitly on the time-dependent parameter $B$ that describes the shape of the Gaussian beam and whose dynamics is governed by the third equation in (\ref{eq:Gaussian_approx}). The remaining two equations describe the dynamics of the overall norm and an additional phase, and can be trivially integrated once the first three equations are solved. In the present paper we shall not be concerned with the dynamics of the phase $\alpha$. 

We can eliminate $p$ to derive a second order differential equation for the position of the wave packet center
\begin{equation}
\ddot{q} = -V_R'(q)+ \frac{\Re(B)}{\Im(B)}V_I'(q)+\frac{1}{\Im(B)} V''_I(q)\dot q,
\end{equation}
that couples to the dynamical equation for the width parameter $B$, and both in turn determine the dynamics of the norm $N$. Since the approximation relies on the Gaussian to be well localized, the imaginary part of $B$ encodes the reliability of the approximation. In what follows we shall consider an example of a model for a $PT$-symmetric waveguide to demonstrate the quality of the approximation for initial Gaussian states, as well as the influence of the beam shape on the propagation. 

\begin{figure}[t]
\centering
\includegraphics[width=0.24\textwidth]{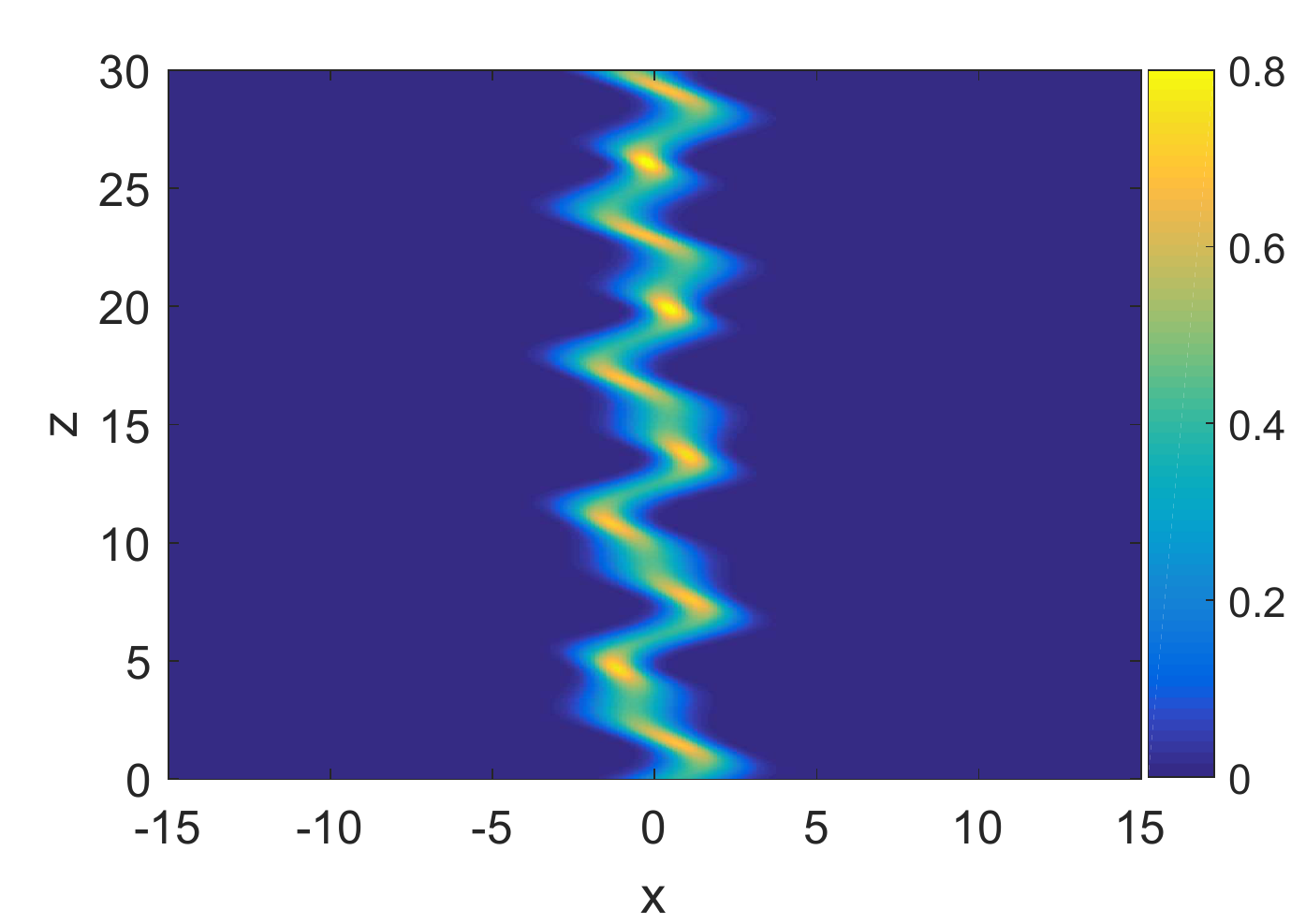}
\includegraphics[width=0.24\textwidth]{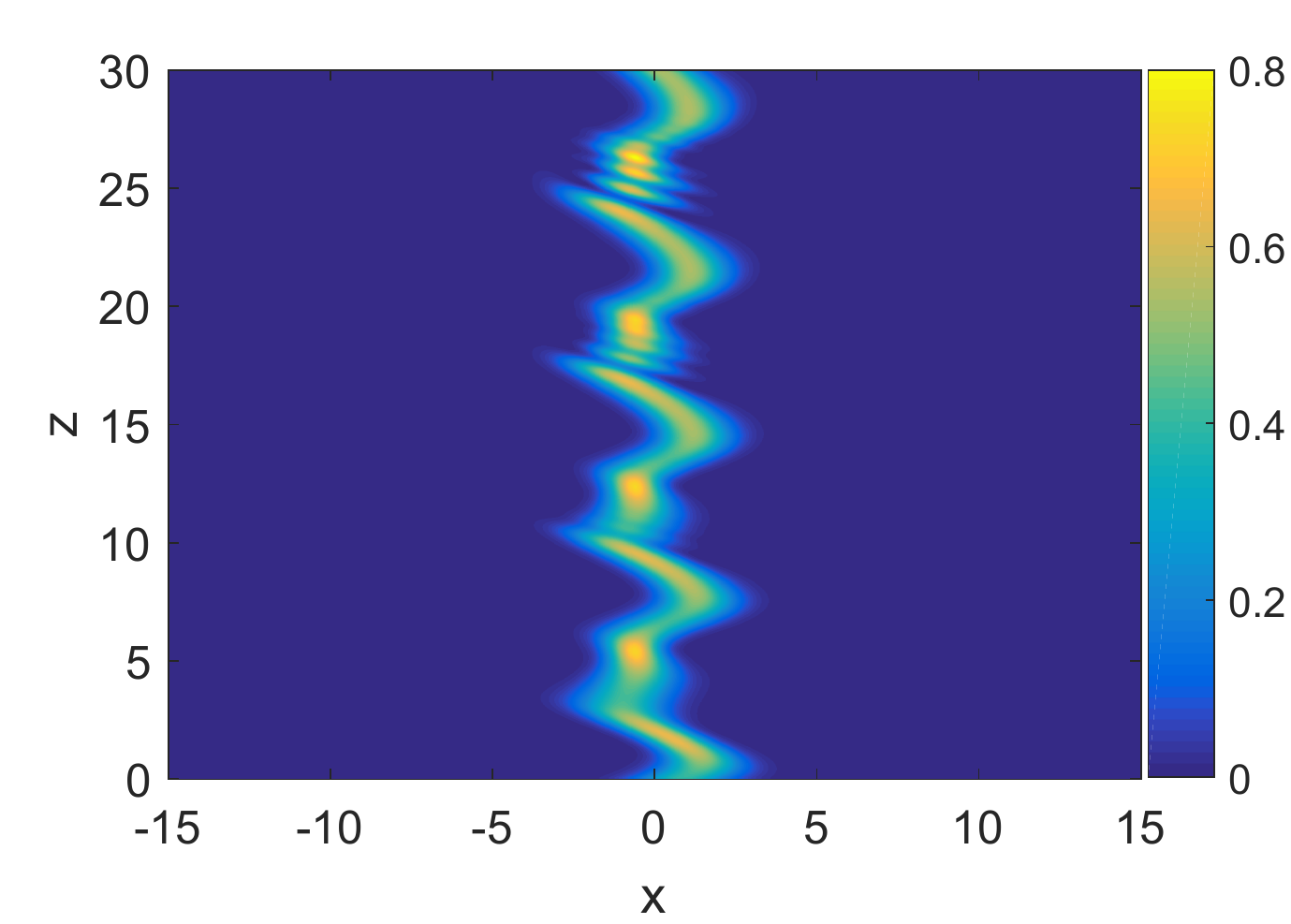}
\includegraphics[width=0.24\textwidth]{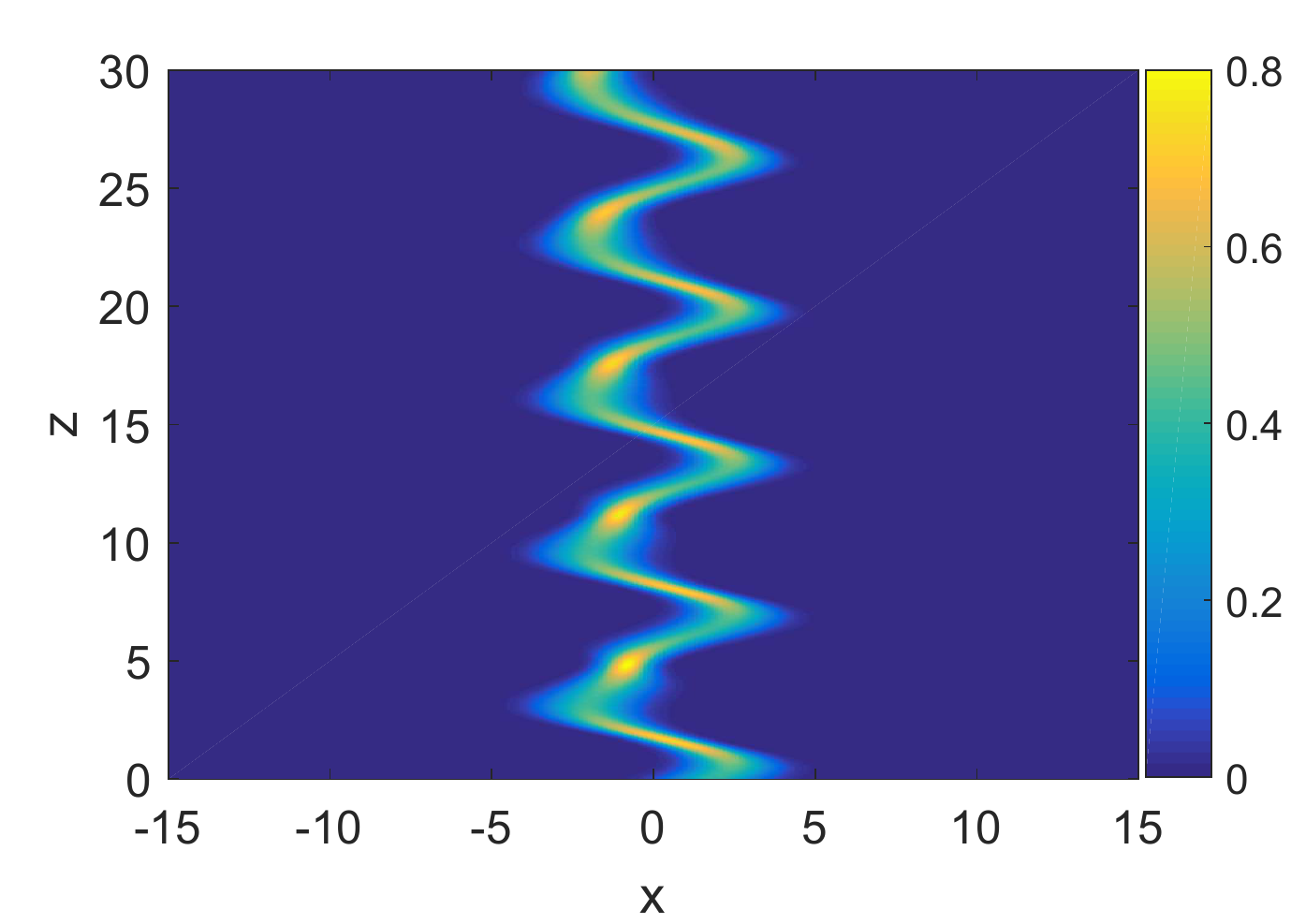}
\includegraphics[width=0.24\textwidth]{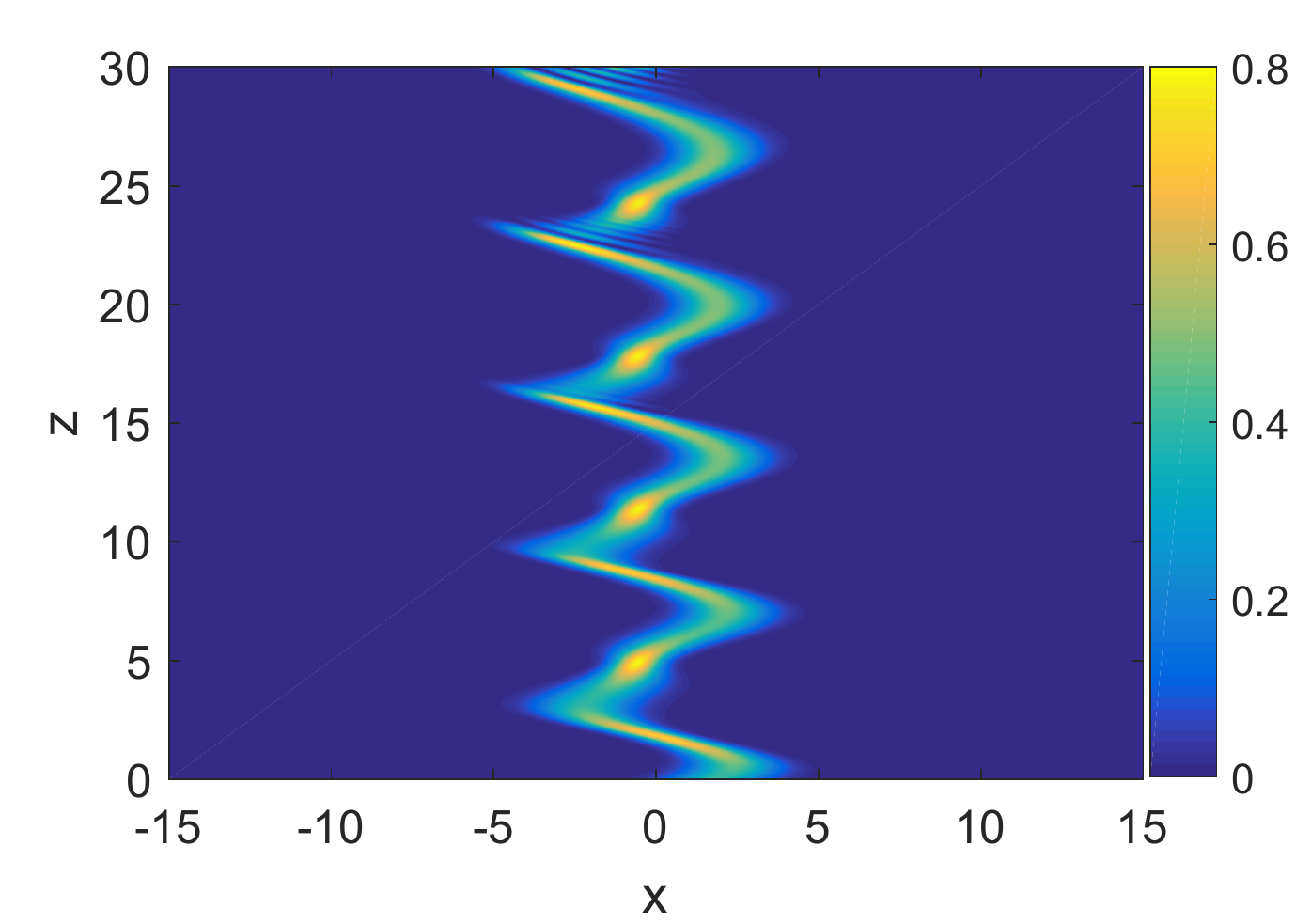}
\includegraphics[width=0.24\textwidth]{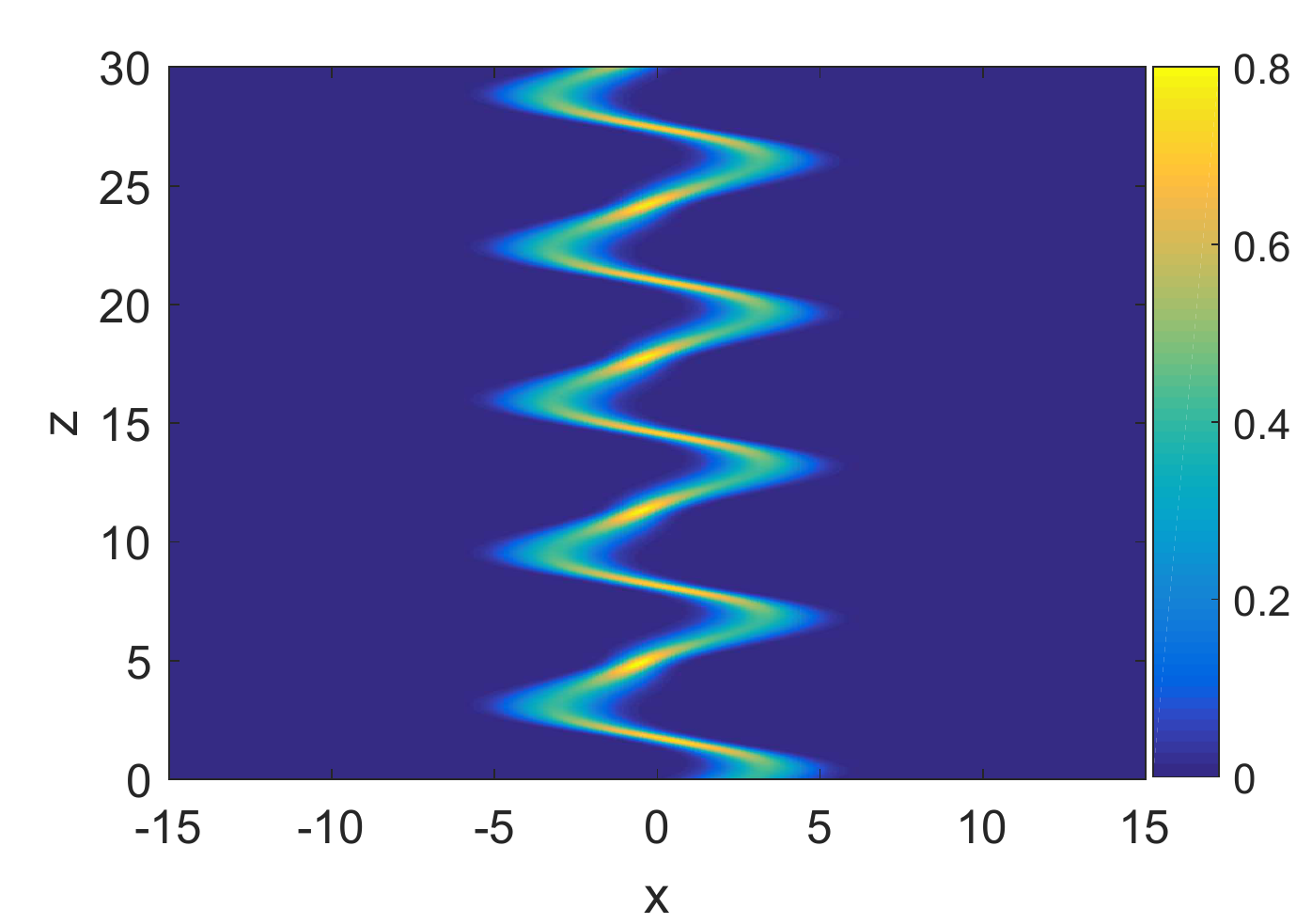}
\includegraphics[width=0.24\textwidth]{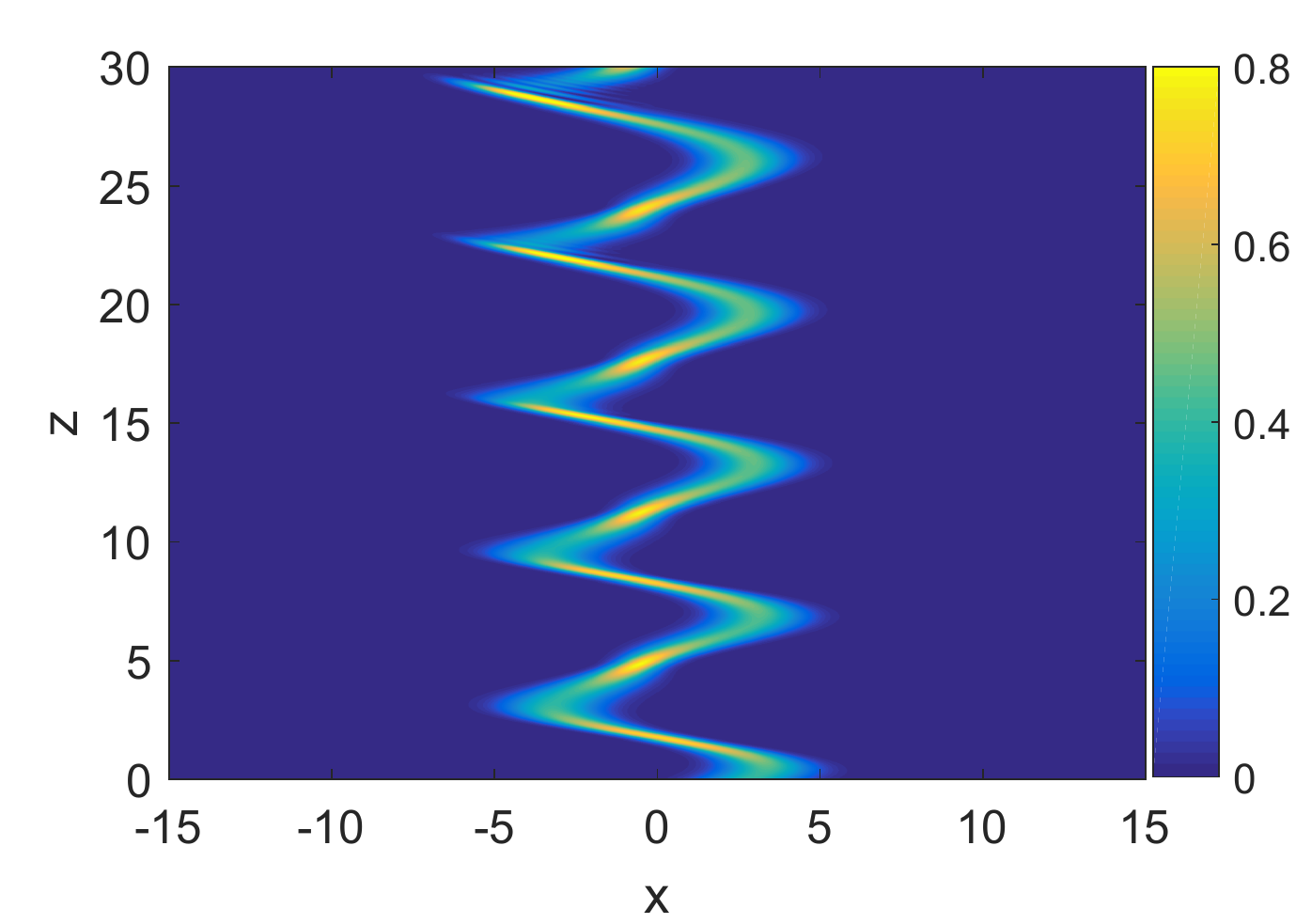}
\includegraphics[width=0.24\textwidth]{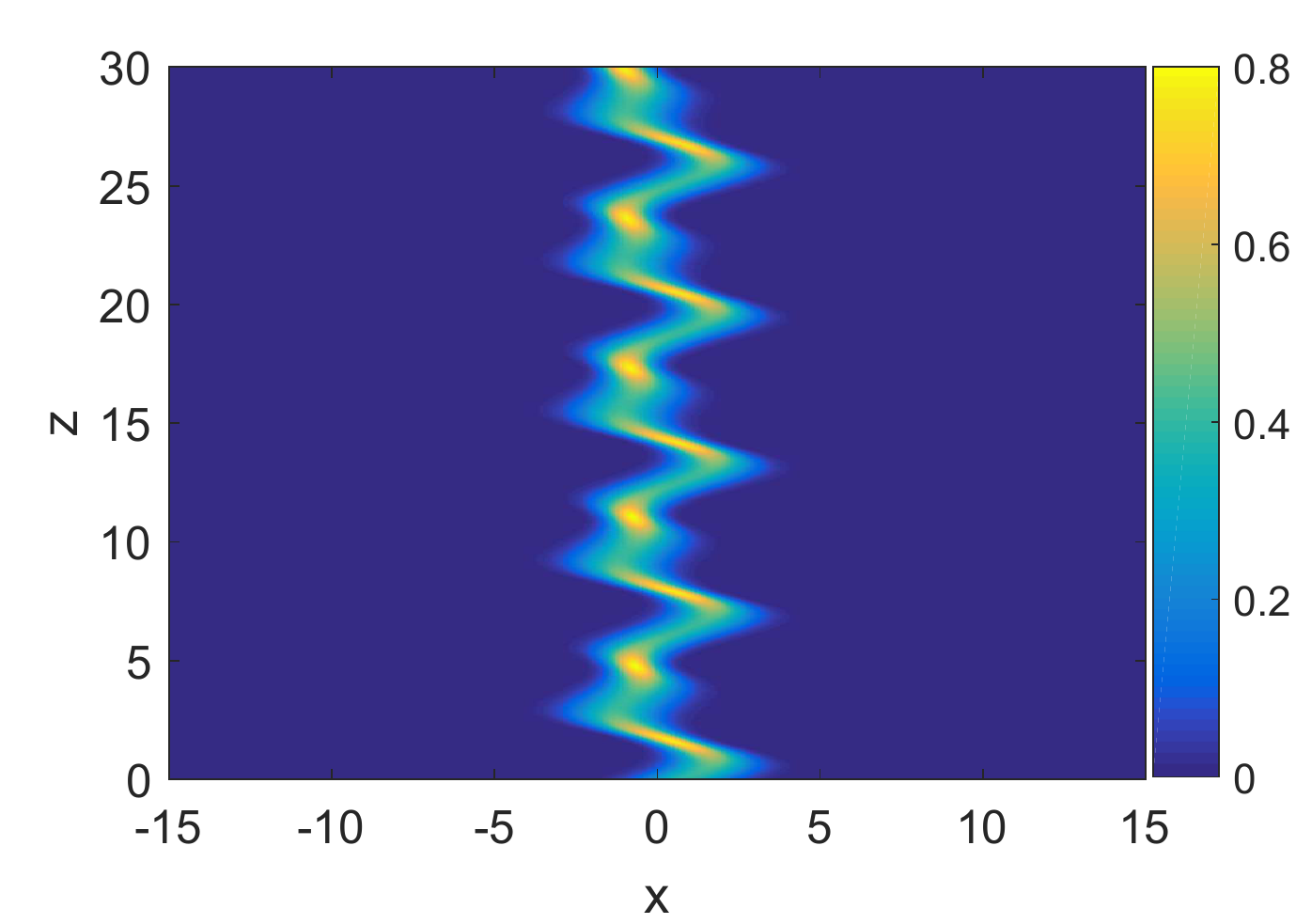}
\includegraphics[width=0.24\textwidth]{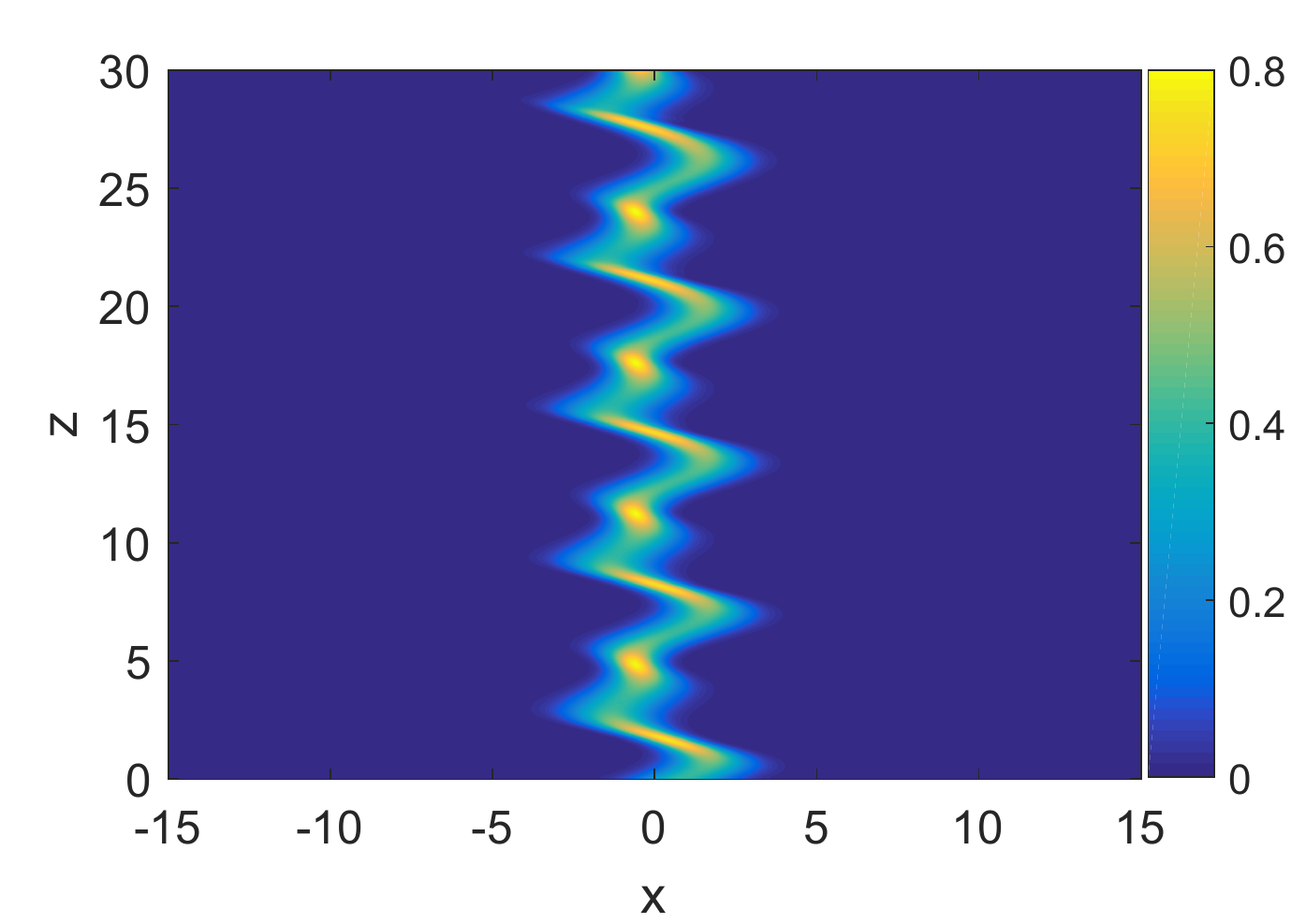}
\caption{Renormalized beam propagation in the Gaussian approximation (left) and for a numerically exact propagation using the split-operator method (right) for $p_0=0$, $B_0=\rmi$ and different values of $q_0$ and $\eta$ ($q_0=1$ and $\eta=5$, $q_0=2$ and $\eta=10$, $q_0=3$ and $\eta=15$, and $q_0=1$ and $\eta=15$ respectively, from top to bottom).}
\label{fig2}
\end{figure}

\begin{figure}[t]
\centering
\includegraphics[width=0.24\textwidth]{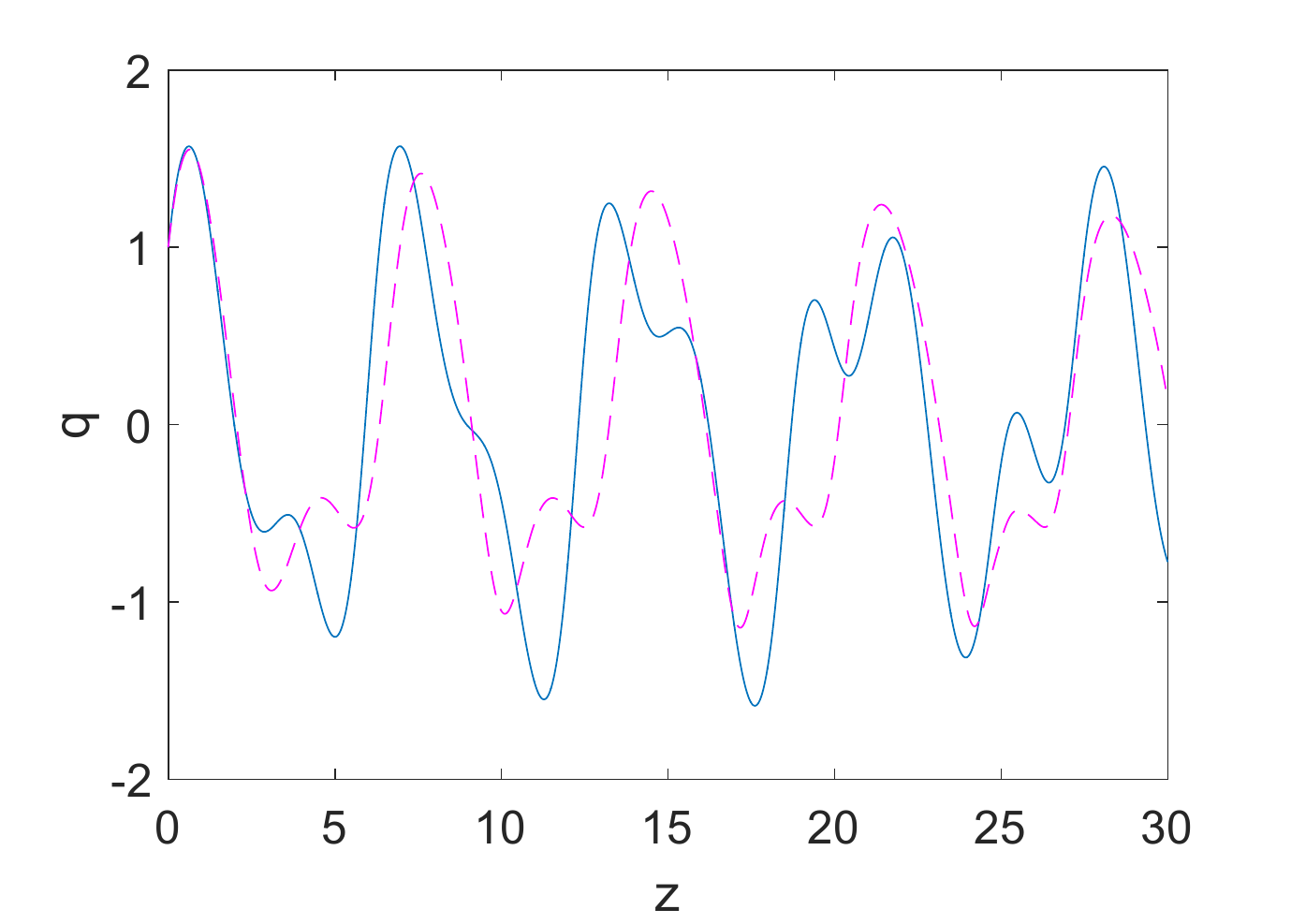}
\includegraphics[width=0.24\textwidth]{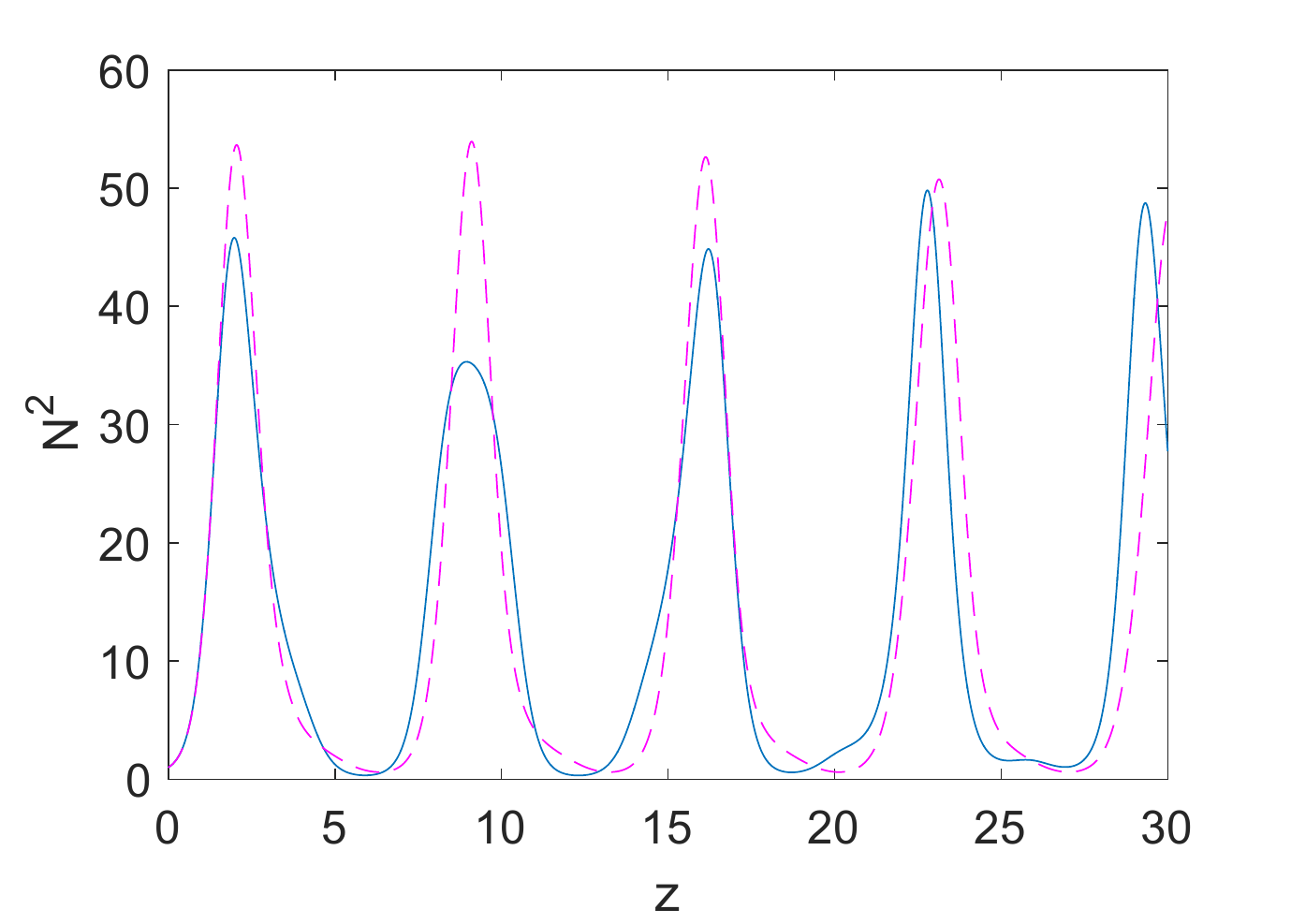}
\includegraphics[width=0.24\textwidth]{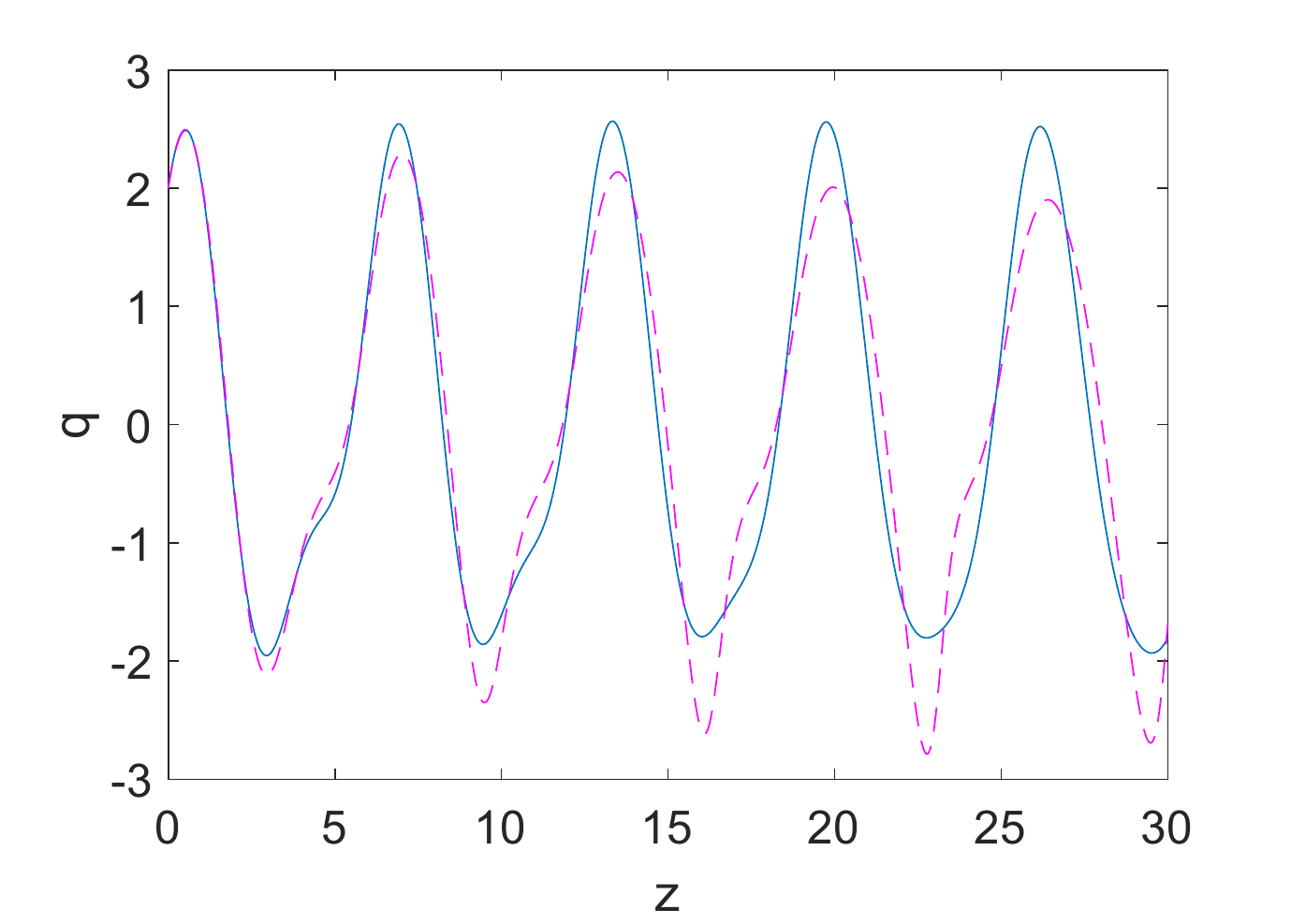}
\includegraphics[width=0.24\textwidth]{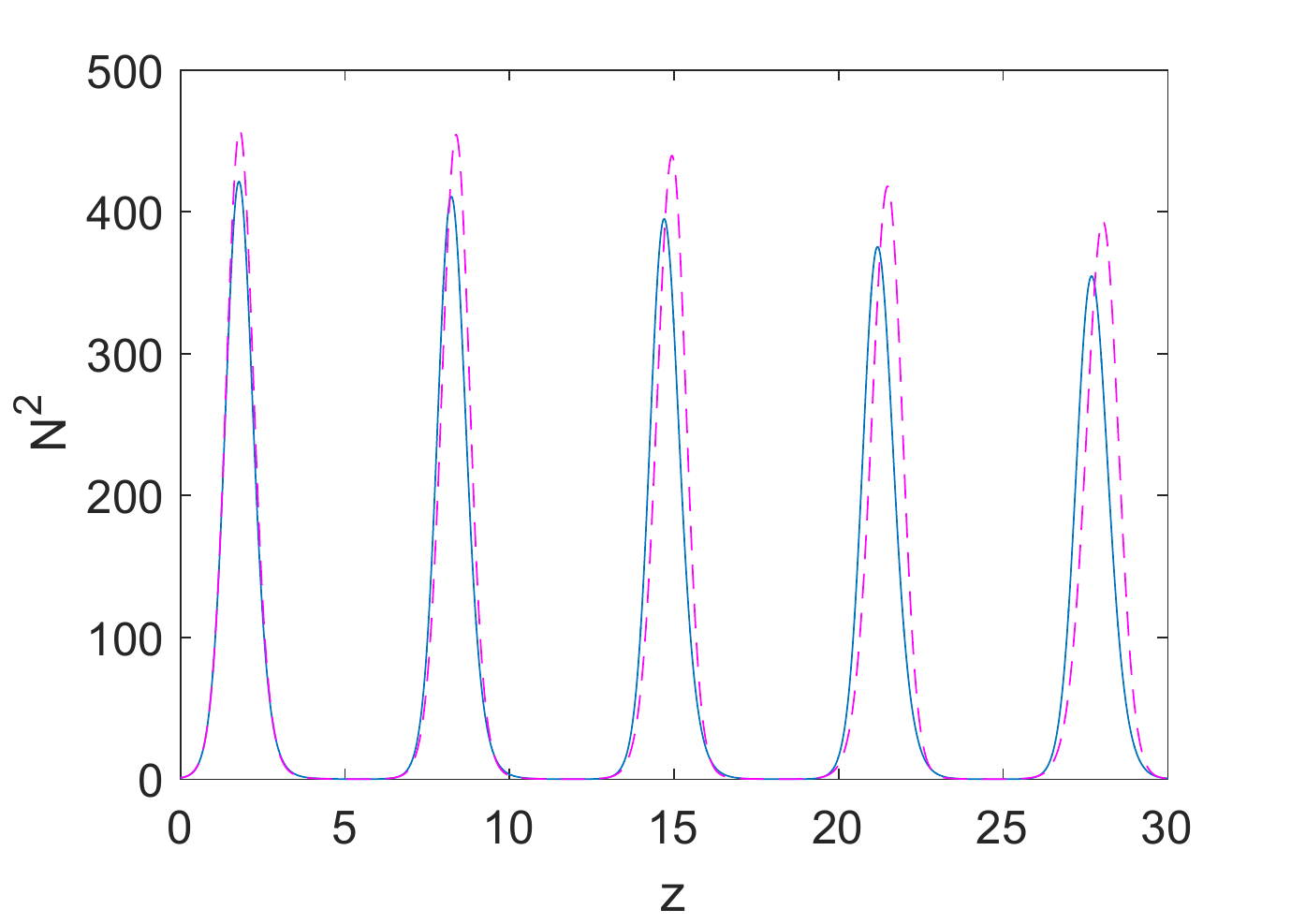}
\includegraphics[width=0.24\textwidth]{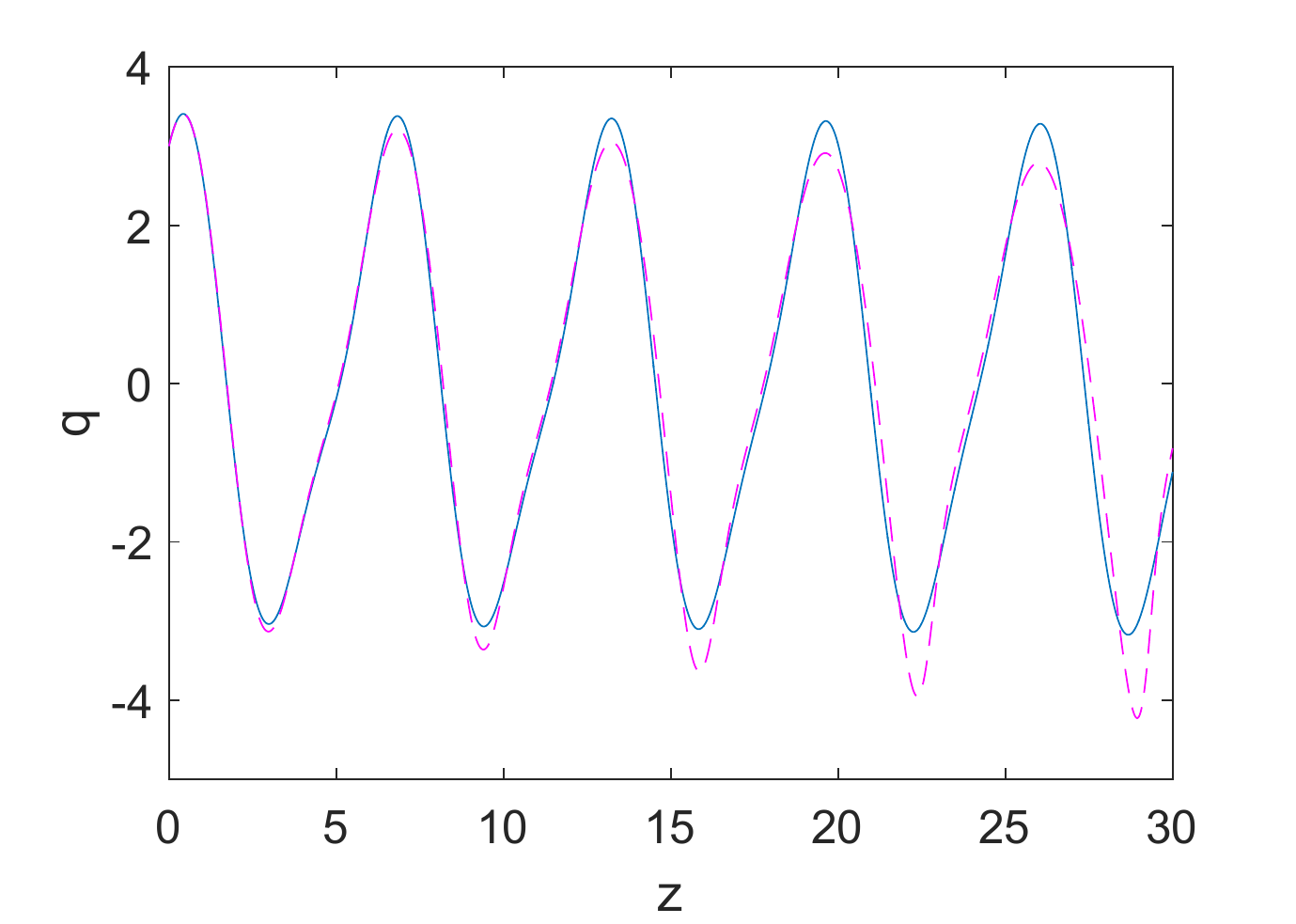}
\includegraphics[width=0.24\textwidth]{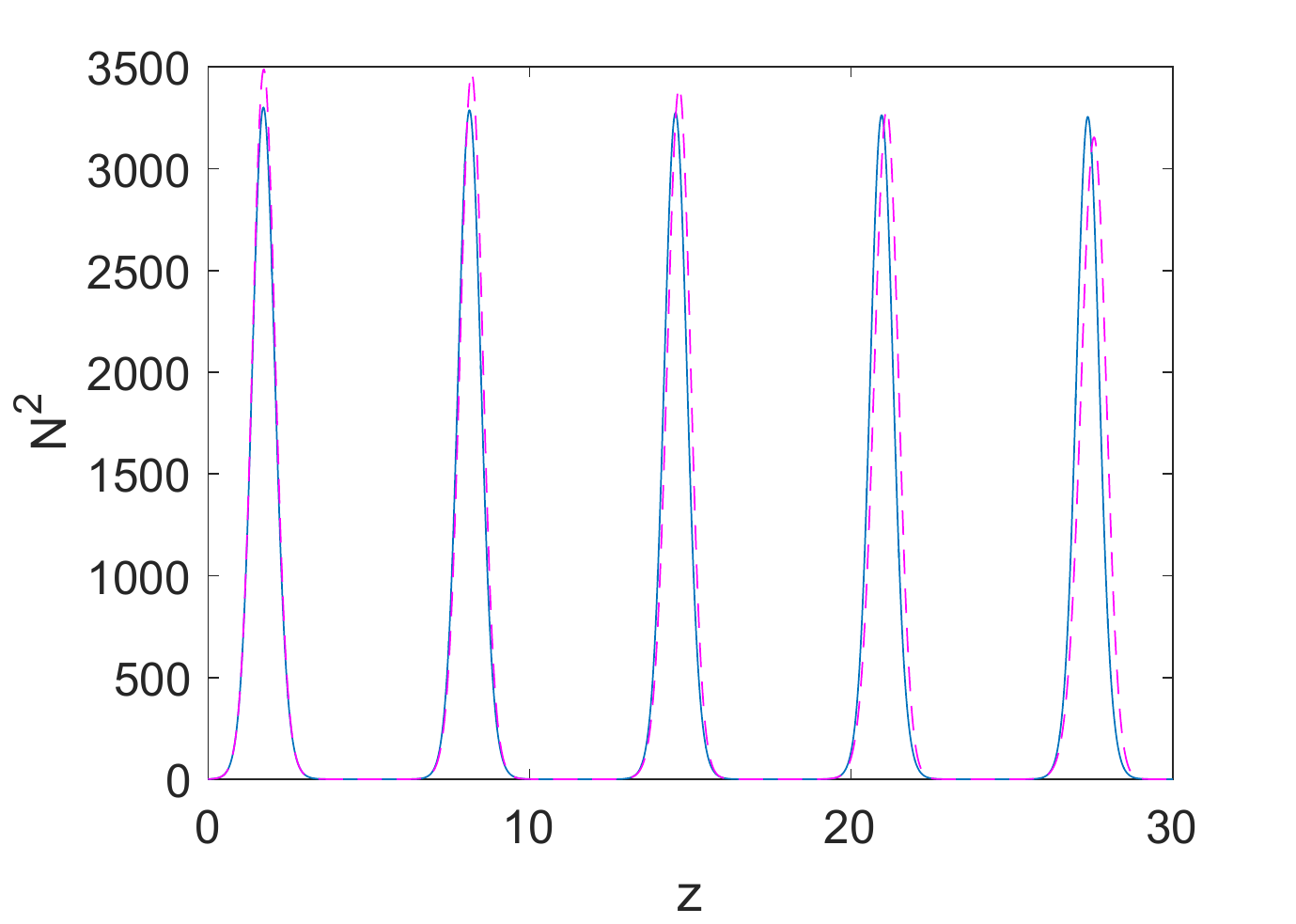}
\includegraphics[width=0.24\textwidth]{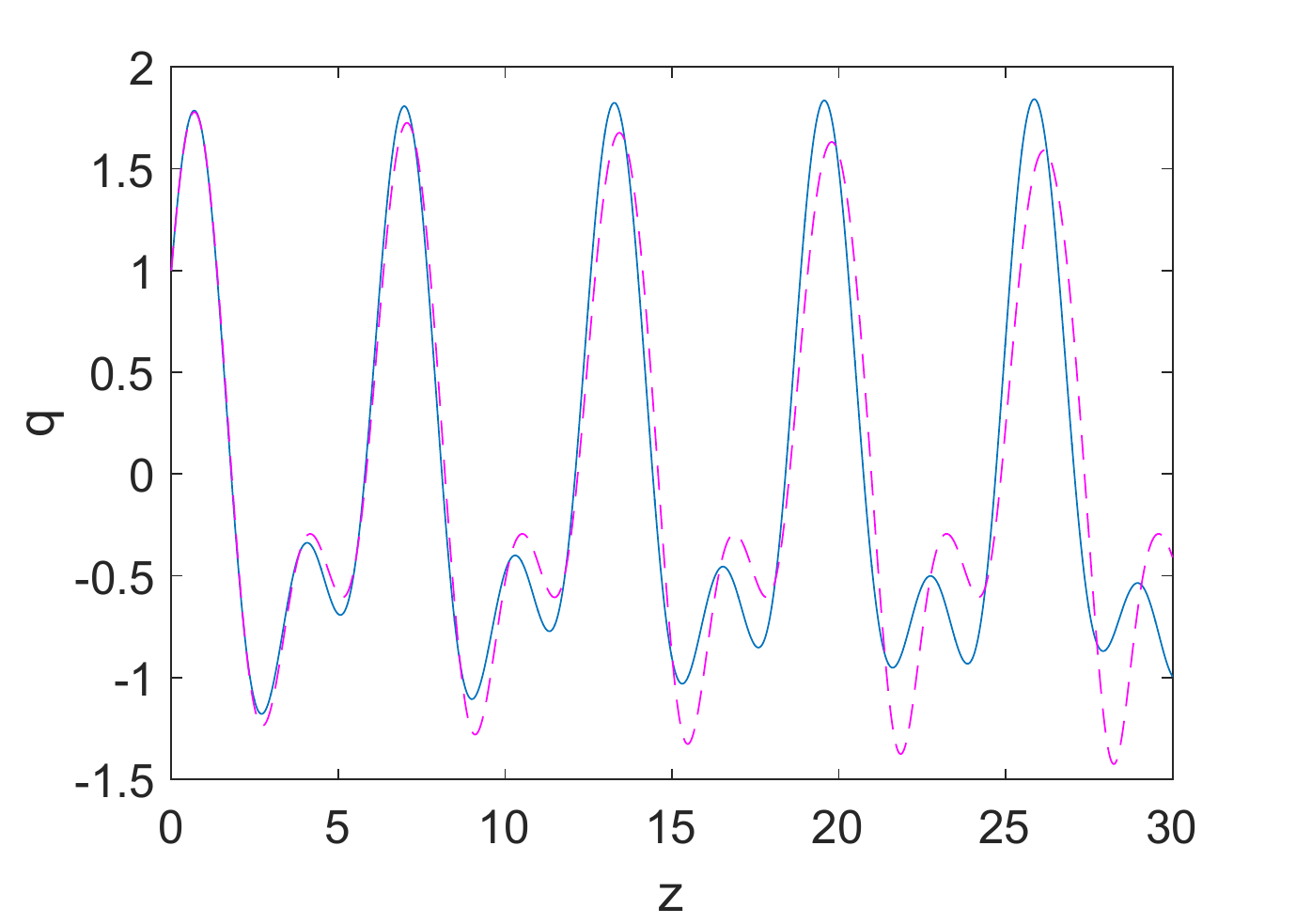}
\includegraphics[width=0.24\textwidth]{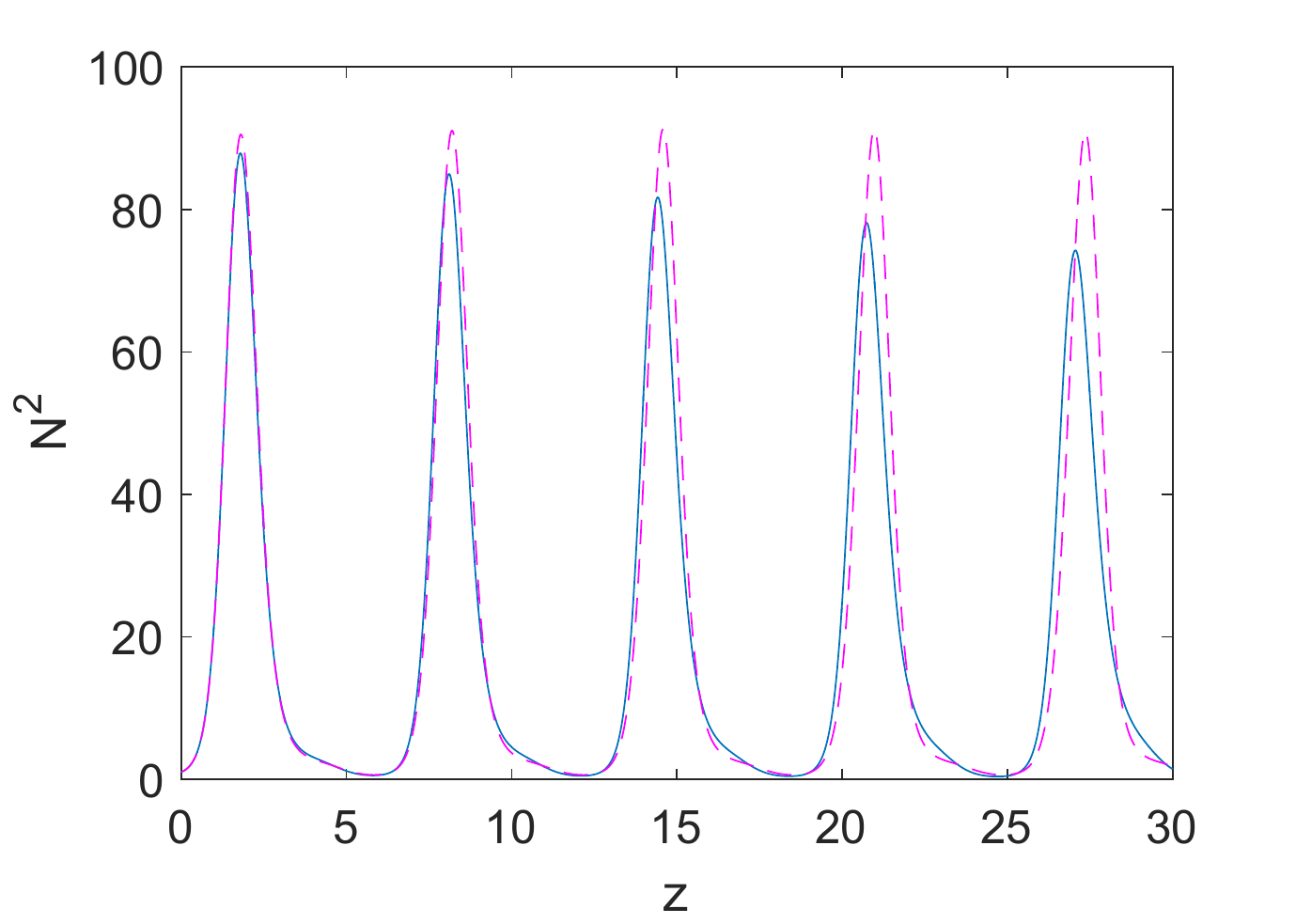}
\caption{Gaussian approximation (solid blue lines) and exact quantum propagation using the split-operator method (dashed magenta lines) for the mean position (left) and the norm (right) for the same parameters as in figure \ref{fig2}.}
\label{fig3}
\end{figure}

\section{Gaussian propagation in a waveguide with $PT$-symmetric gain-loss profile}
We consider the complex potential
\begin{equation}
\label{eq:pot_ex1}
V = - \left(1- \rmi\frac{\gamma}{\eta}\tanh\left(\frac{x}{\eta}\right)\right)\eta^2 \rme^{-\frac{\omega^2x^2}{2\eta^2}}
\end{equation}
as a model for a multimode waveguide with $PT$-symmetric gain-loss profile. The parameter $\eta$ determines the scale of the potential, which extends further and is deeper for larger values of $\eta$. Thus, the larger the value of $\eta$, the more classical (i.e. described by equations (\ref{eq:Gaussian_approx})) we expect the propagation of Gaussian states to be. The strength of the loss-gain profile is described by the parameter $\gamma$. Close to the origin the real part of the potential is approximately harmonic with frequency $\omega$ independently of the value of $\eta$. Thus, the fundamental mode of the real part of the potential is approximately a Gaussian state with $B=\rmi\omega$, i.e., with width $\Delta q=\frac{1}{\sqrt{2\omega}}$, which provides the natural length scale on which the potential should be locally well described by the second order Taylor expansion for the Gaussian approximation to be valid. Figure \ref{fig_pot1} depicts the real and imaginary part of the potential for $\gamma=1$ and $\omega=1$ for two values of $\eta$. For comparison the intensity profile for a Gaussian beam with $B=\rmi$ is also depicted. 

To illustrate how the Gaussian approximation improves with increasing $\eta$ we show examples of the propagation in the approximation as well as the numerically exact solution of the Schr\"odinger equation (using a split-operator method) in Figure \ref{fig2} for three different values of $\eta$ for $B_0=\frac{\rmi}{2}$, $p_0=0$ and initial positions slightly off the central axis. We have chosen a value of the width that is different from that of the coherent state of the harmonic approximation around the origin, such that there is a stronger contribution from the width. Since the potential is not completely scale-invariant with respect to $\eta$ we cannot directly compare the propagation for different values of $\eta$ and corresponding initial values of $q_0$. To provide an overview we show three examples with different values of $\eta$ for the same value of $\frac{q_0}{\eta}$ as well as a fourth example with a large value of $\eta$ where the initial value of $q_0$ is chosen the same as for the example with the smallest value of $\eta$. Figure \ref{fig3} depicts the corresponding propagation of the mean values of the position and the norm. It can be observed that while the Gaussian approximation becomes more accurate for larger values of $\eta$ as expected, it already qualitatively captures characteristic features of the evolution even for relatively small values of $\eta$. We shall return to a more detailed analysis and explanation of some of these features later. In the following we shall focus on the case $\eta=10$, where we expect the Gaussian approximation to make reasonable predictions for propagation distances of the order of a few oscillations. 

\begin{figure}[t]
\centering
\includegraphics[width=0.24\textwidth]{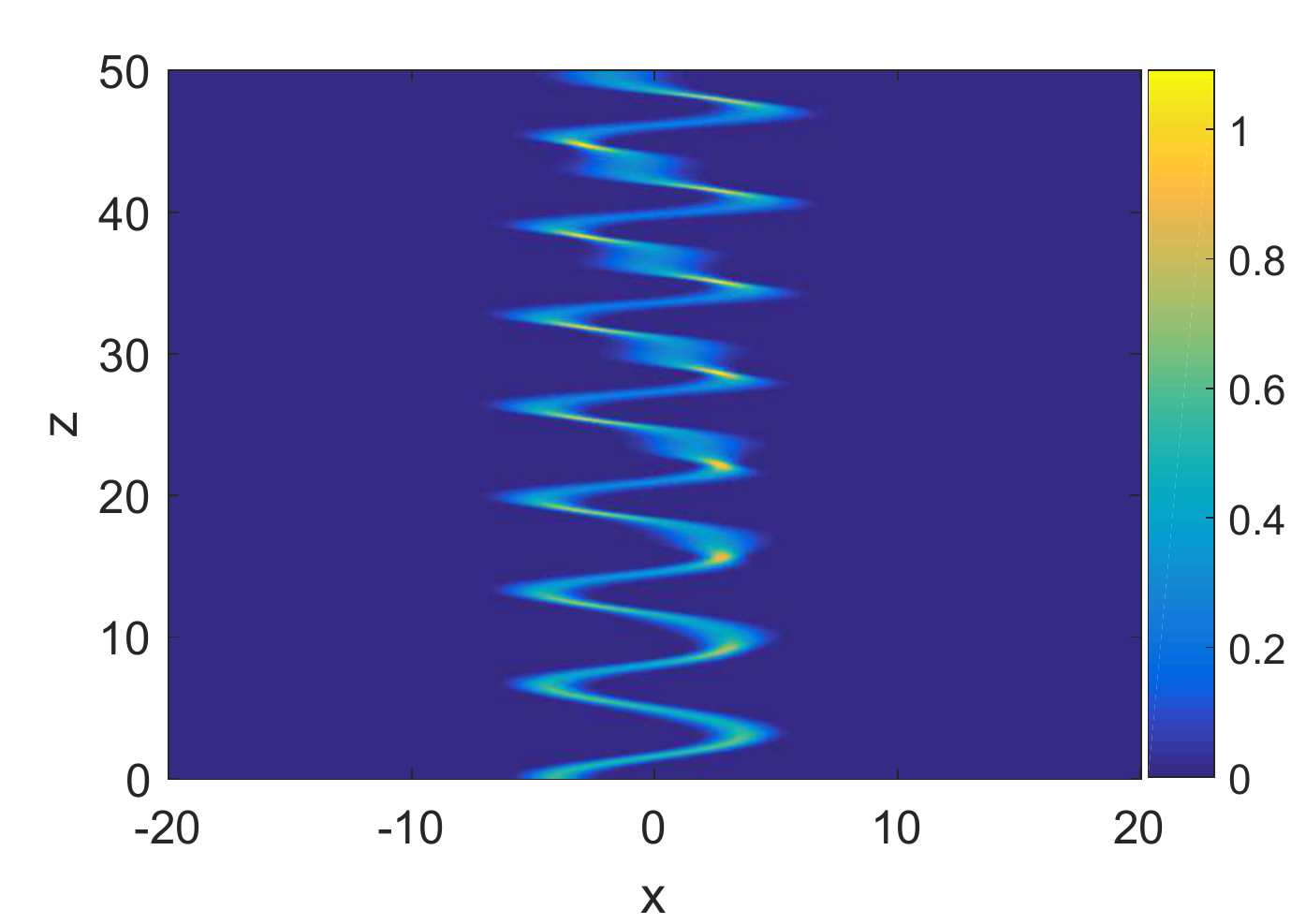}
\includegraphics[width=0.24\textwidth]{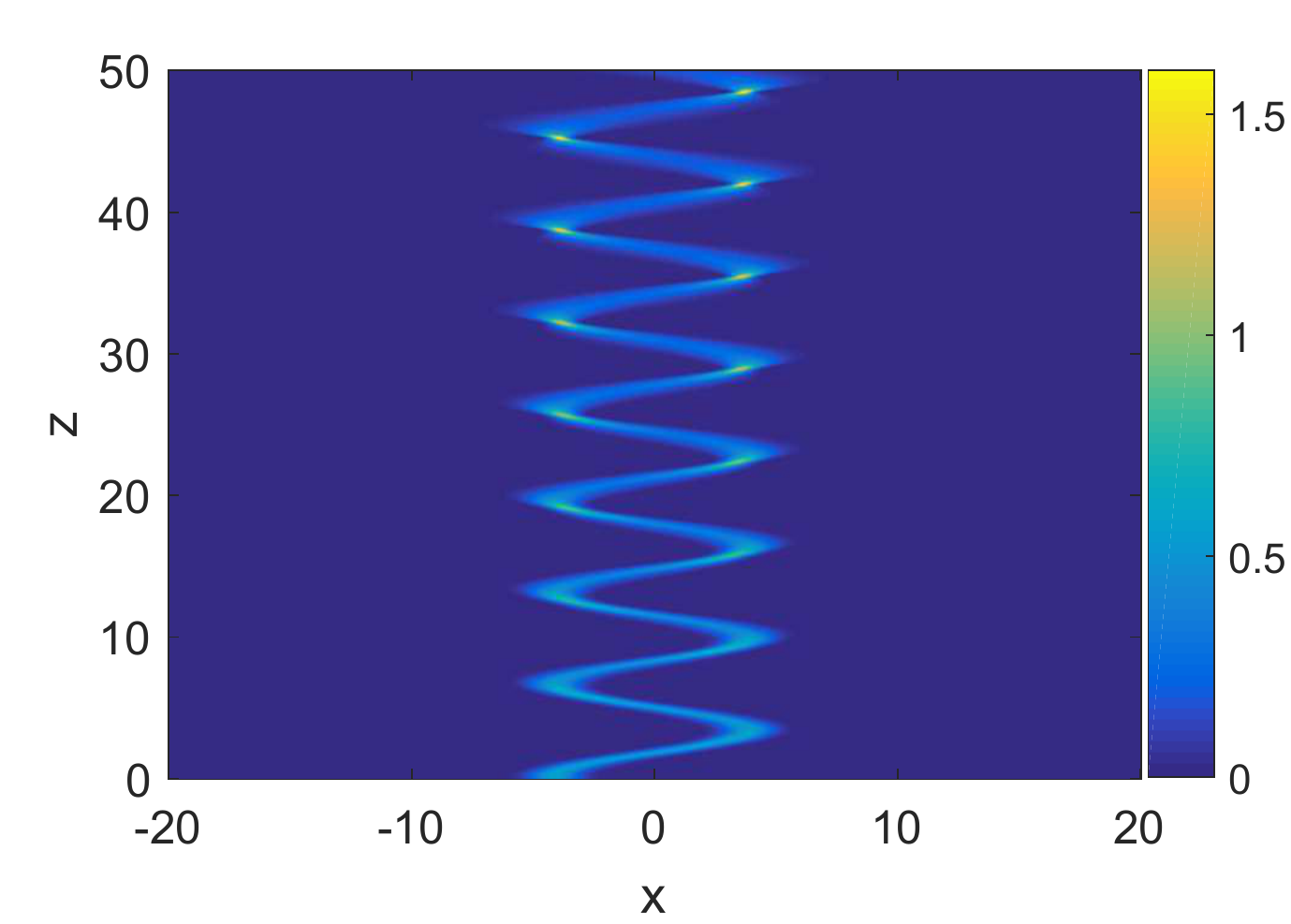}
\includegraphics[width=0.24\textwidth]{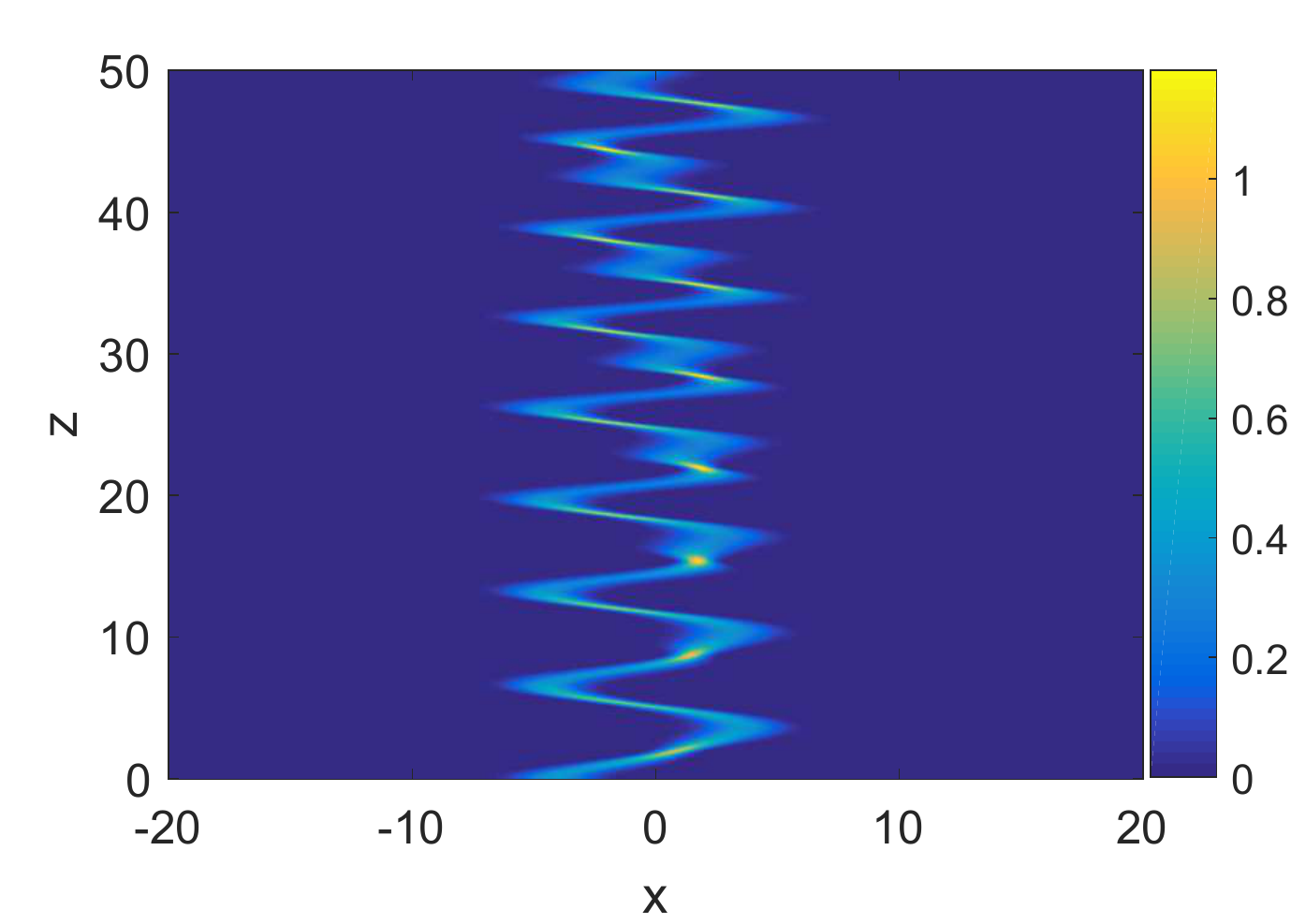}
\includegraphics[width=0.24\textwidth]{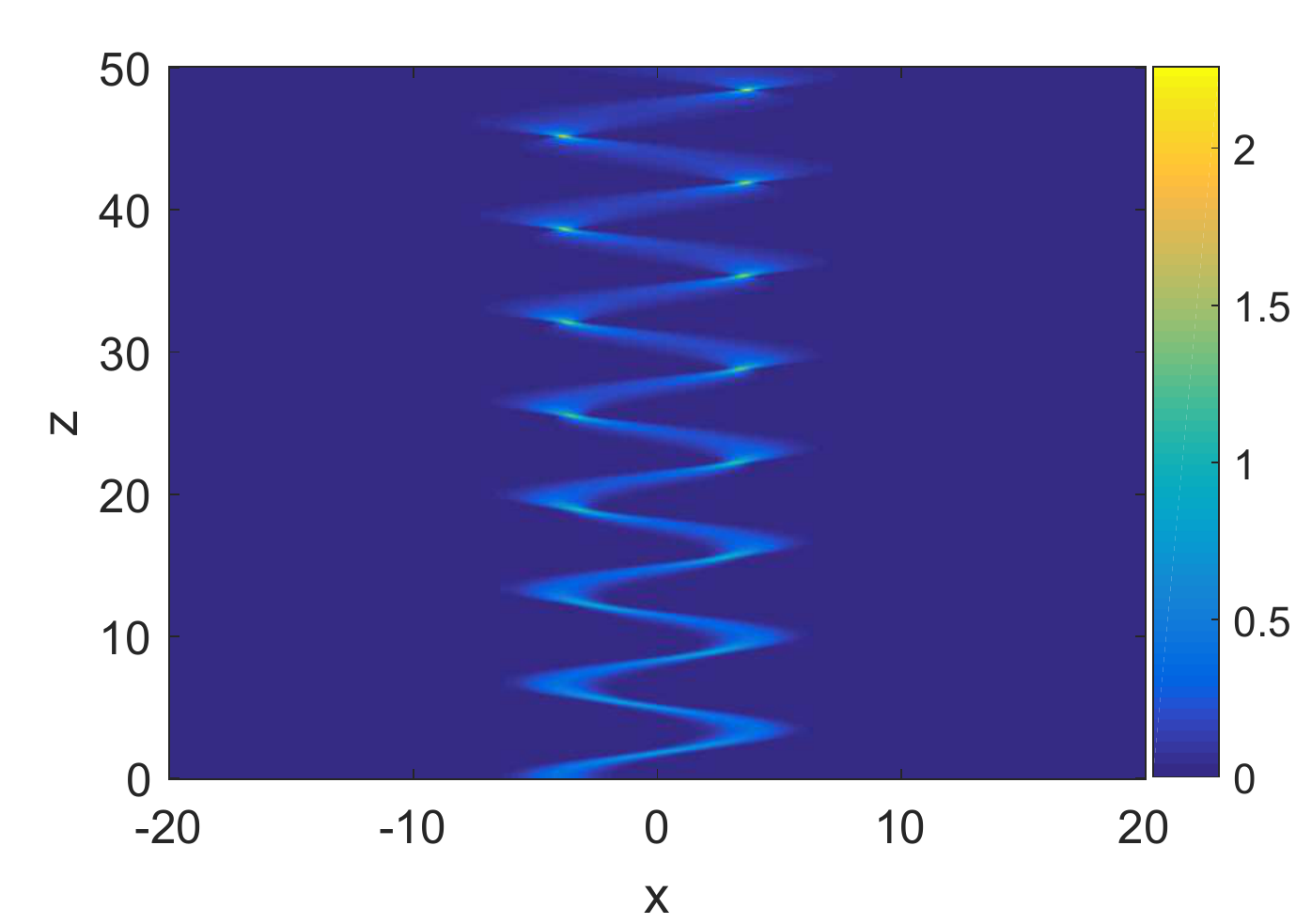}
\includegraphics[width=0.24\textwidth]{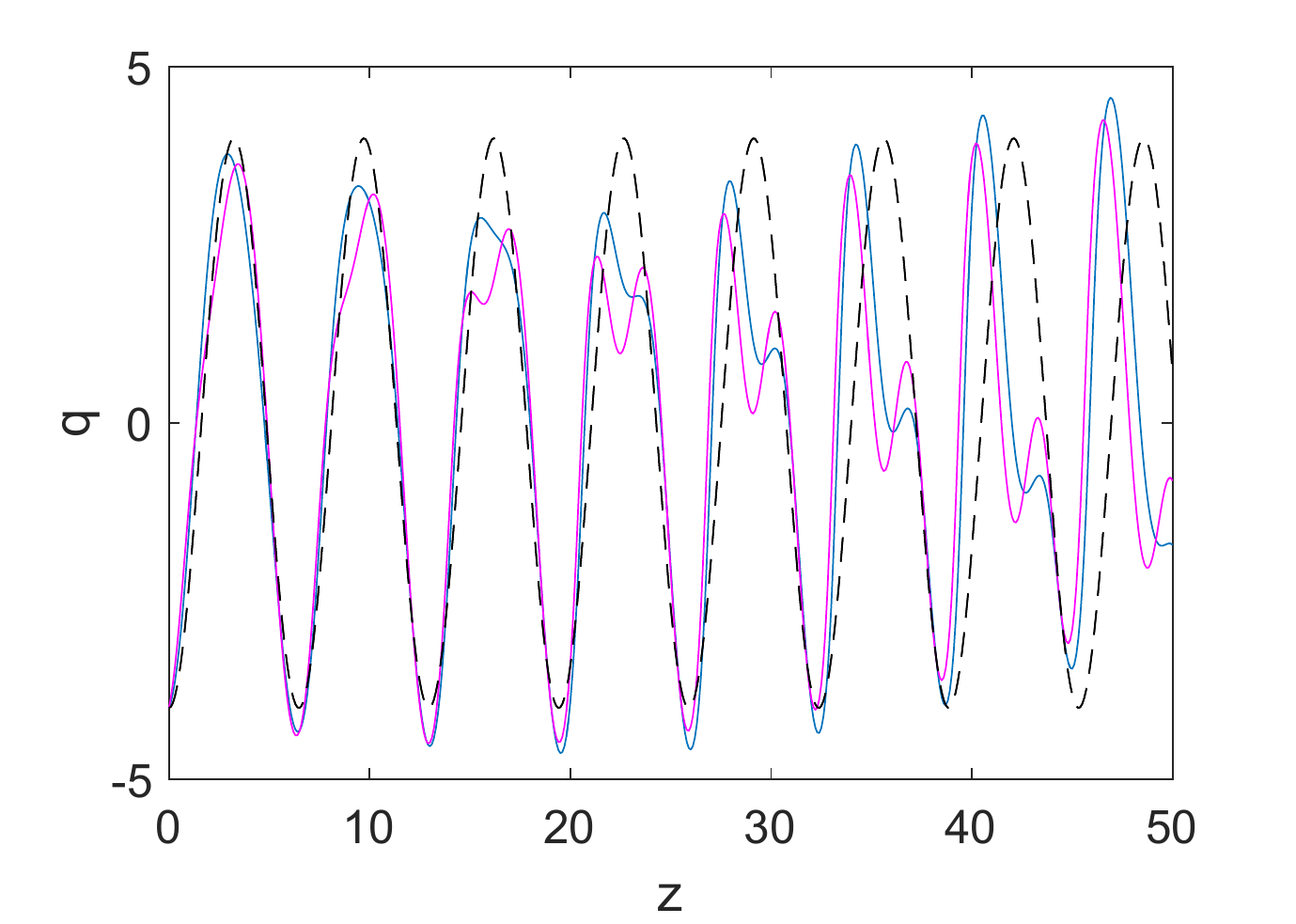}
\includegraphics[width=0.24\textwidth]{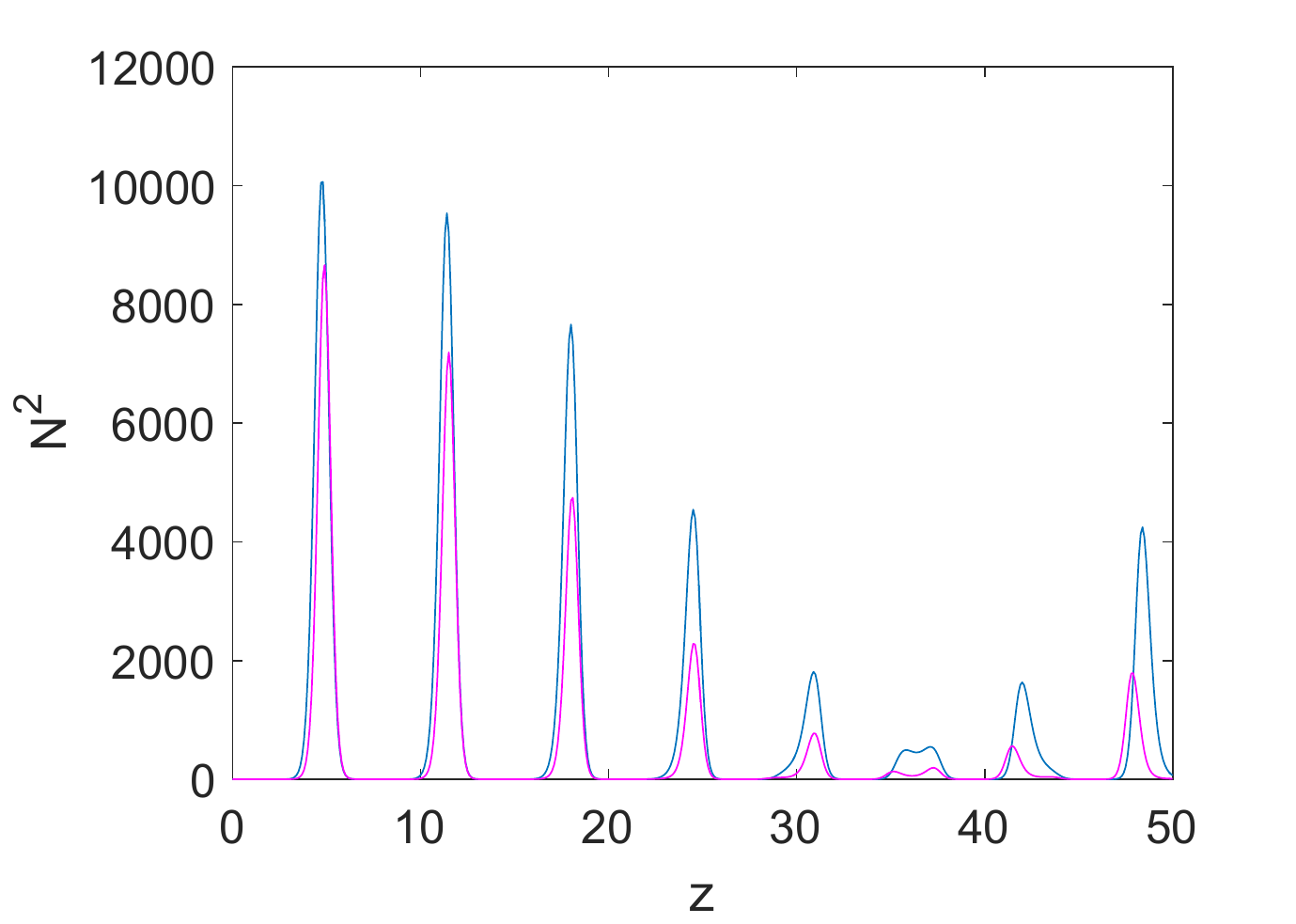}
\caption{The top row shows the (renormalised) propagation in the Gaussian approximation for $\eta=10$, $B_0=\rmi$, $p_0=0$, and $q_0=-4$ on the left in comparison to the Gaussian approximation for the equivalent Hermitian system on the right. The second row shows the same propagations for $B_0=\frac{\rmi}{2}$. The bottom left figure shows the comparison between the mean position in the presence (solid blue line for $B_0=\rmi$, solid magenta line for $B_0=\frac{\rmi}{2}$) and absence (dashed black line) of gain and loss, corresponding to the figures in the top two rows. The right figure on the bottom shows the norm of the propagated beam in the presence of gain and loss for $B_0=\rmi$ (blue) and $B_0=\frac{\rmi}{2}$ (magenta).}
\label{fig4}
\end{figure}

Figure \ref{fig4} shows the Gaussian approximation for the propagation of initial Gaussian beams with $B_0=\rmi$ as well as $B_0=\frac{\rmi}{2}$ and a relatively large initial displacement from the central axis of $q_0=-4$ (i.e. in the loss region) in comparison to the propagation of the same initial beam in the absence of loss and gain, i.e. for $\gamma=0$. The corresponding mean positions are depicted in direct comparison in the lower left panel of the same figure. The dominant feature in the beam propagation even in the presence of gain and loss in the current example are the oscillations due to the real part of the potential. We note a reduction of the symmetry and modulations on top of the periodic oscillations induced by the loss-gain profile, partly related to the fact that the varying width of the wave packet influences the central motion. The norm of the propagated beam depicted in the lower panel on the right, however, is strongly influenced by the gain and loss, and shows pronounced modulations as the beam oscillates between the loss and the gain regions. The difference between the two initial widths is only a quantitative one here. We have found the Gaussian approximation to be in good correspondence with the numerically exact propagation for the propagation distance depicted here.
\begin{figure}[t]
\centering
\includegraphics[width=0.24\textwidth]{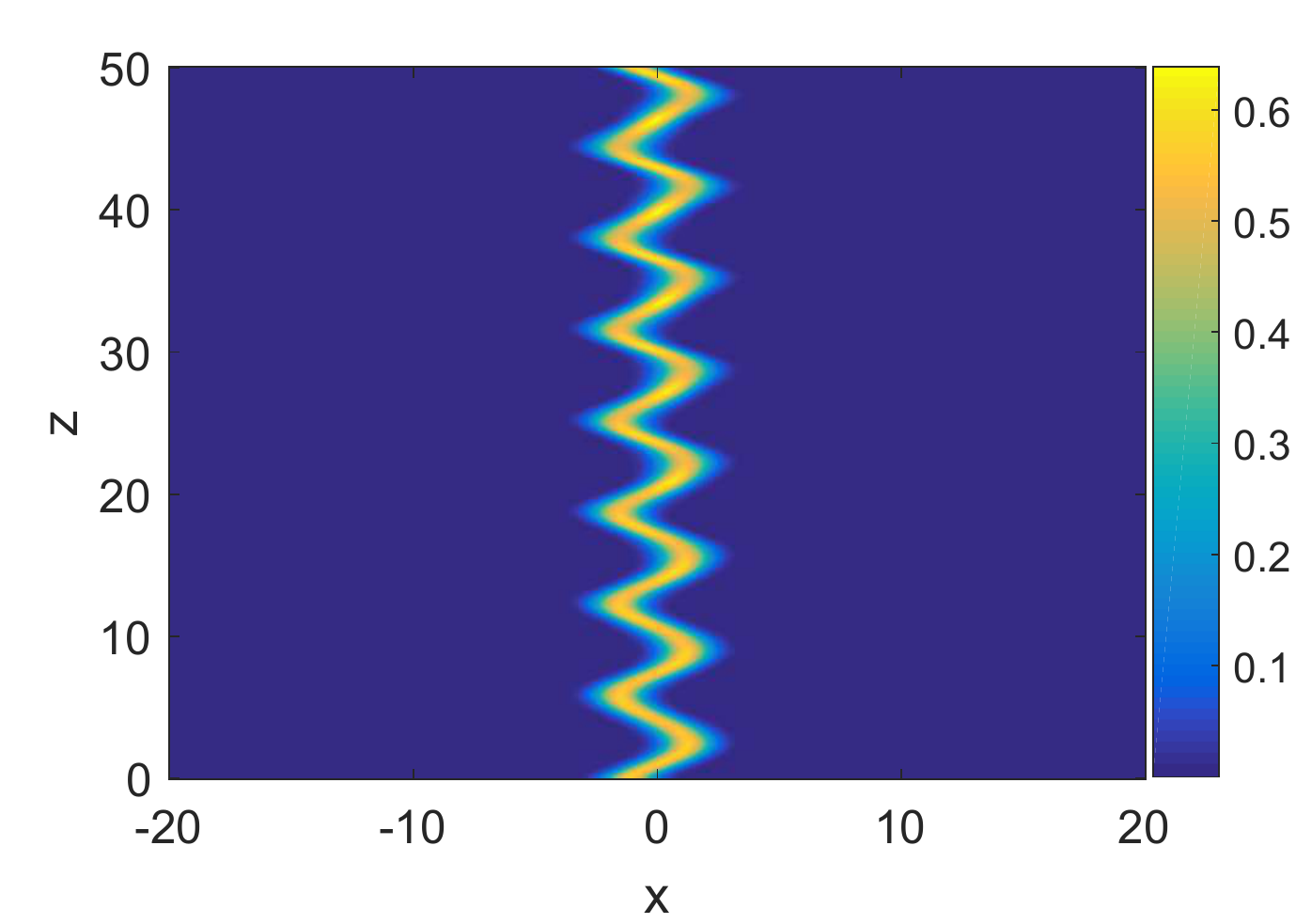}
\includegraphics[width=0.24\textwidth]{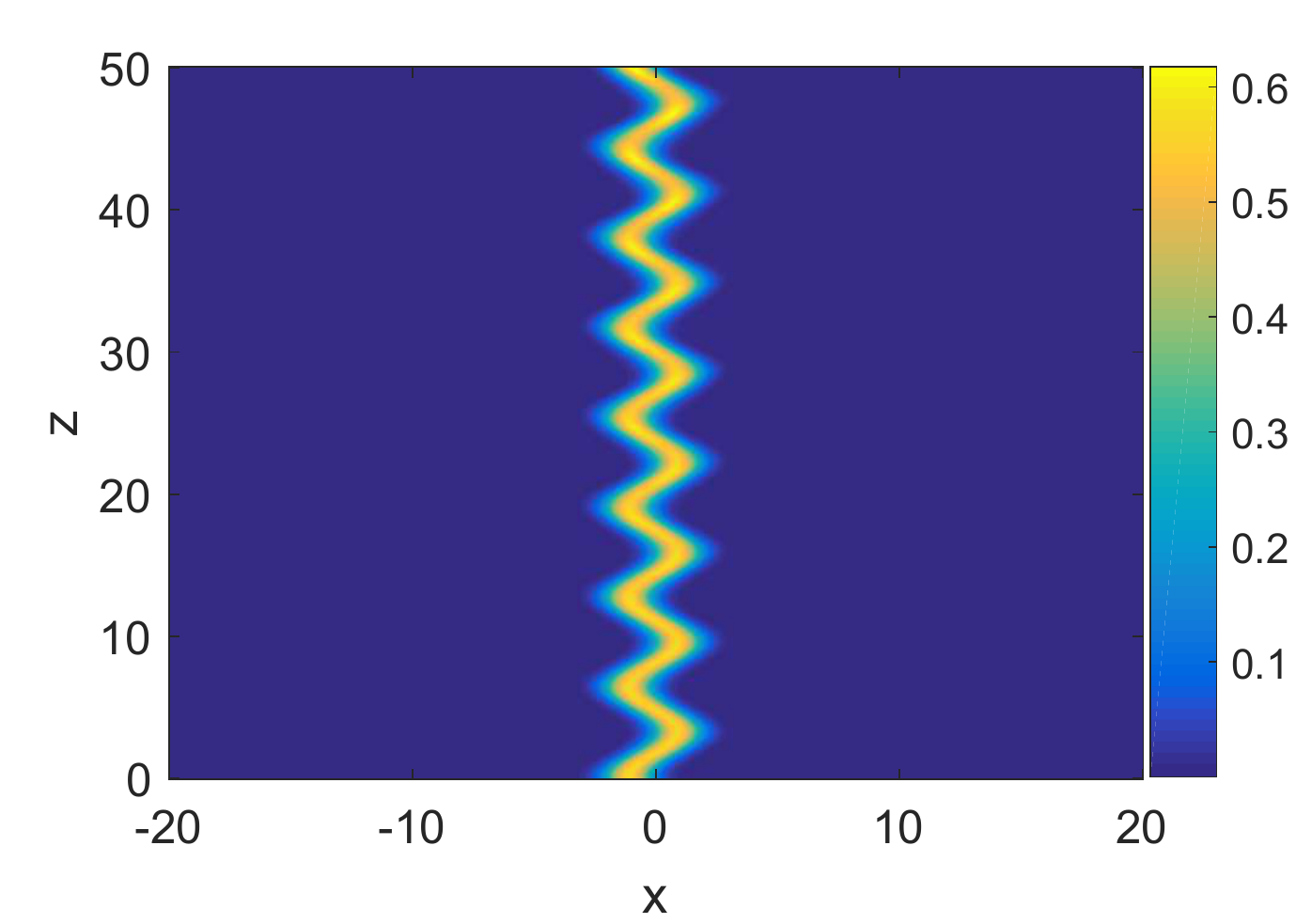}
\includegraphics[width=0.24\textwidth]{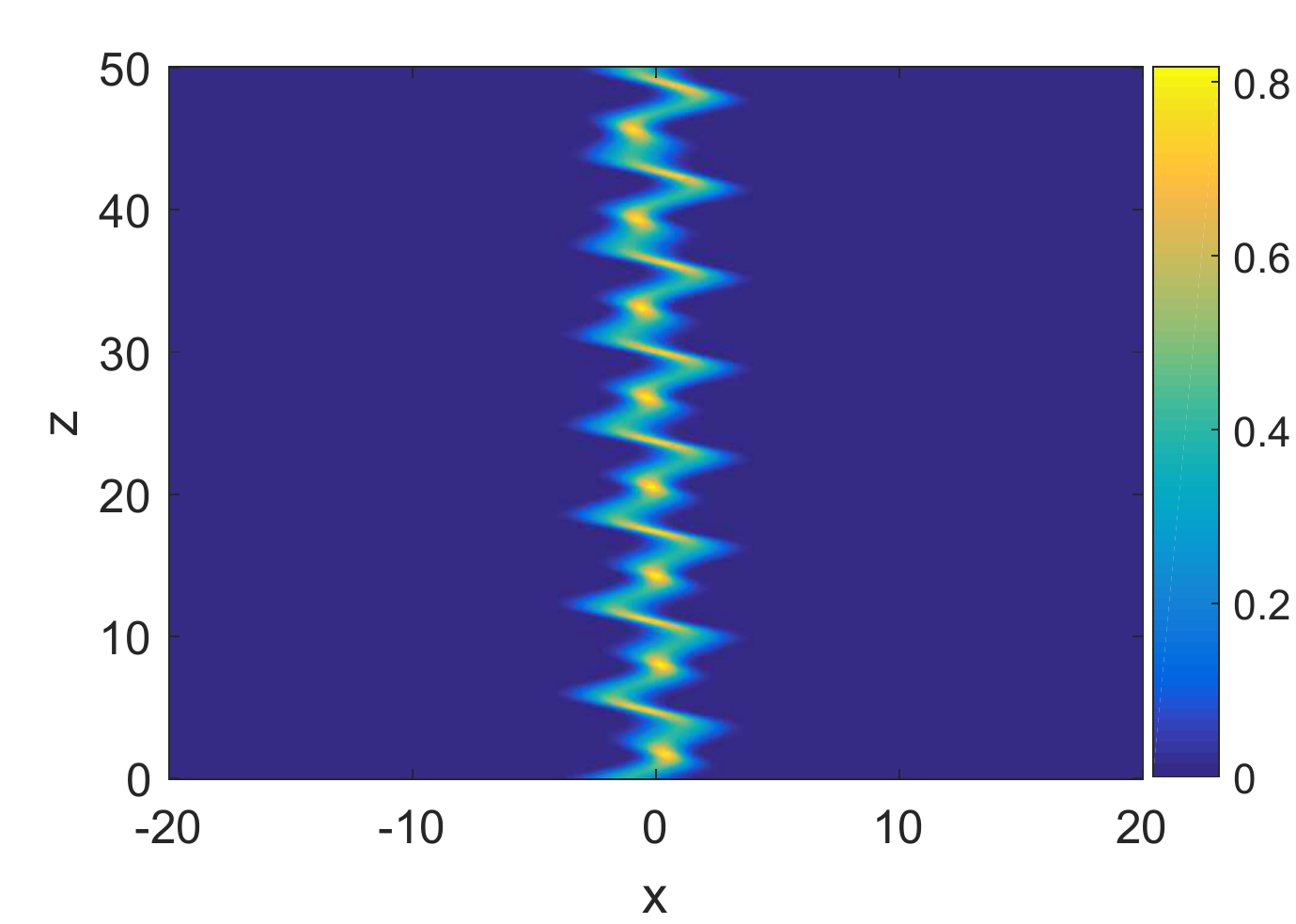}
\includegraphics[width=0.24\textwidth]{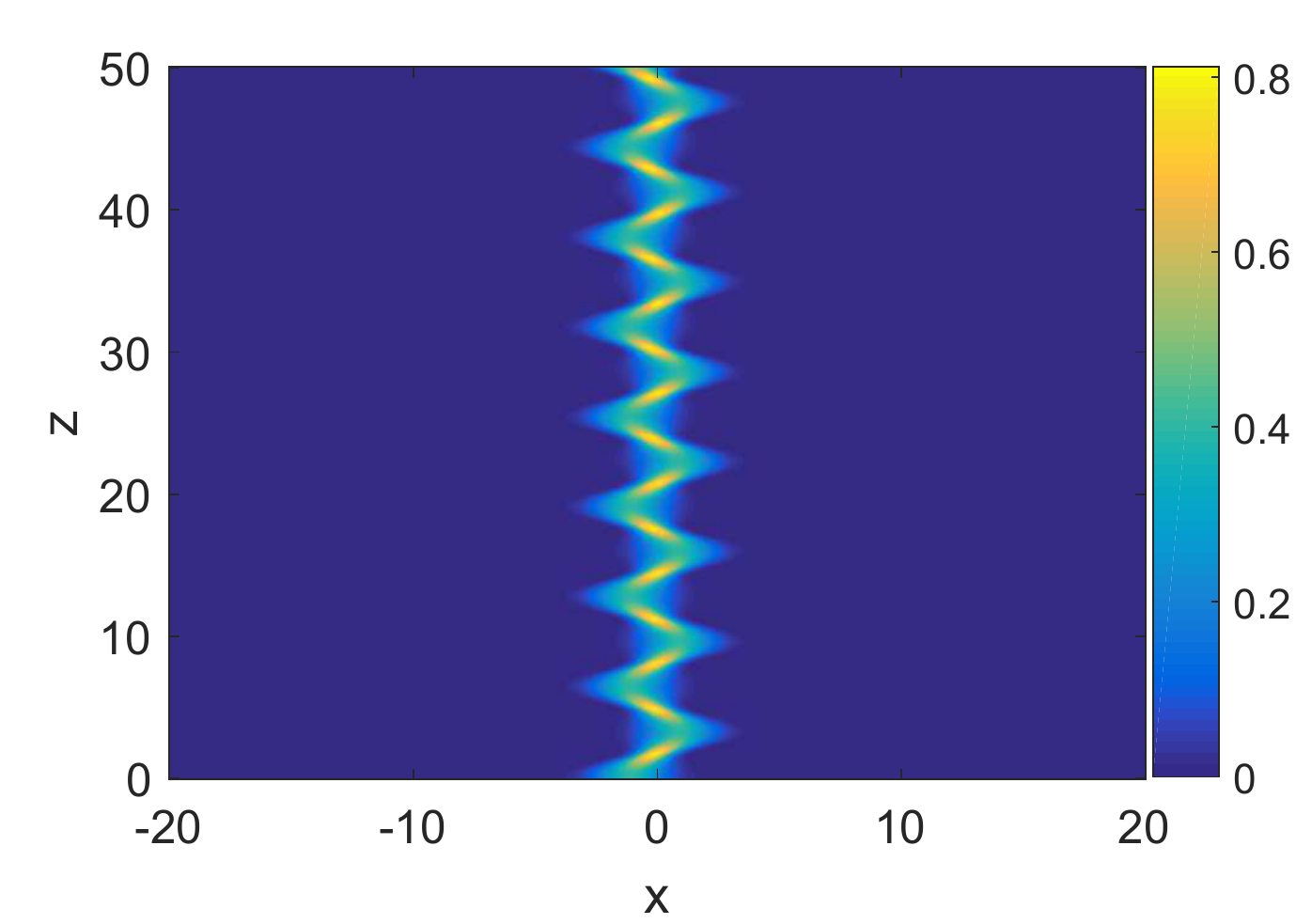}
\includegraphics[width=0.24\textwidth]{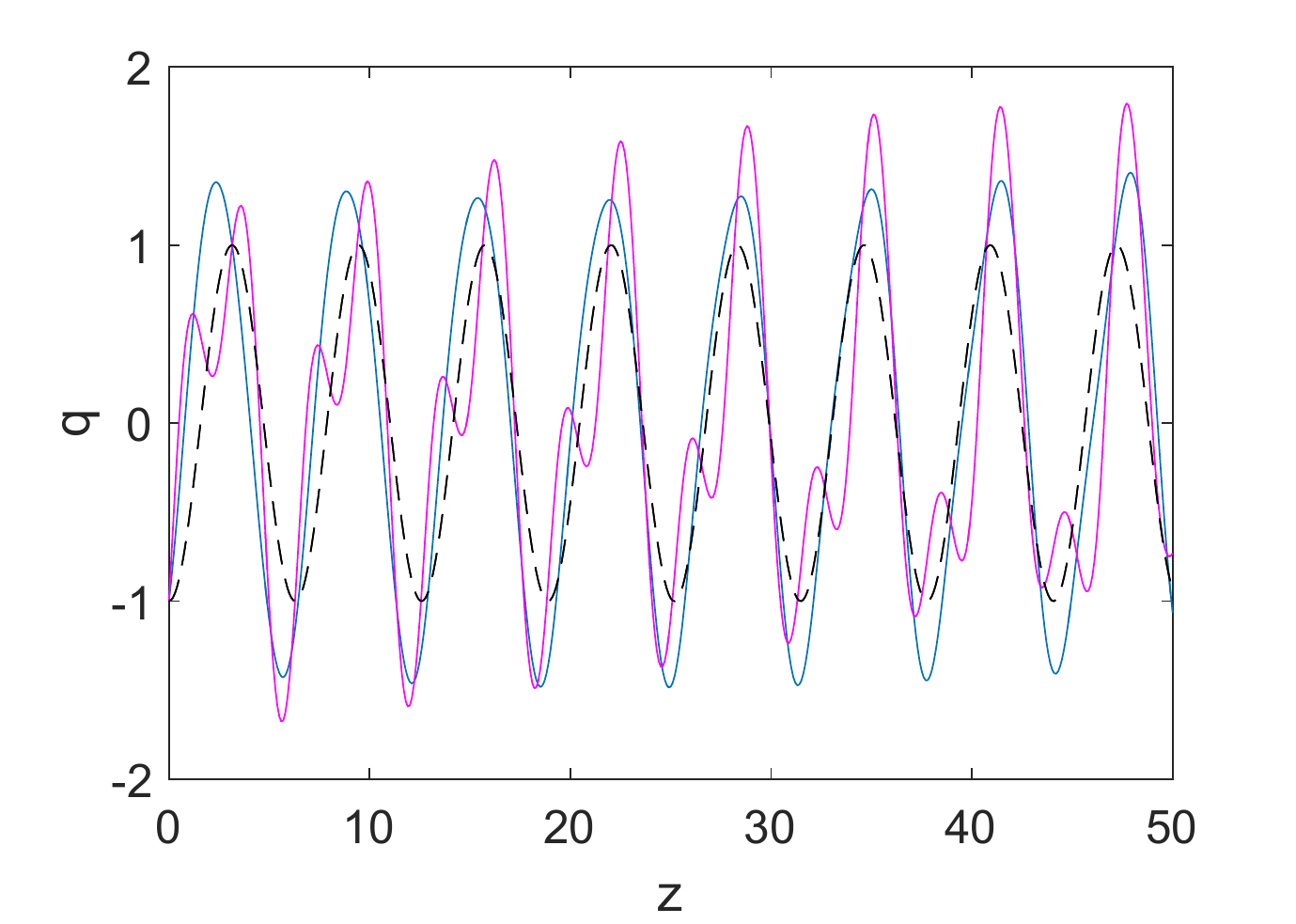}
\includegraphics[width=0.24\textwidth]{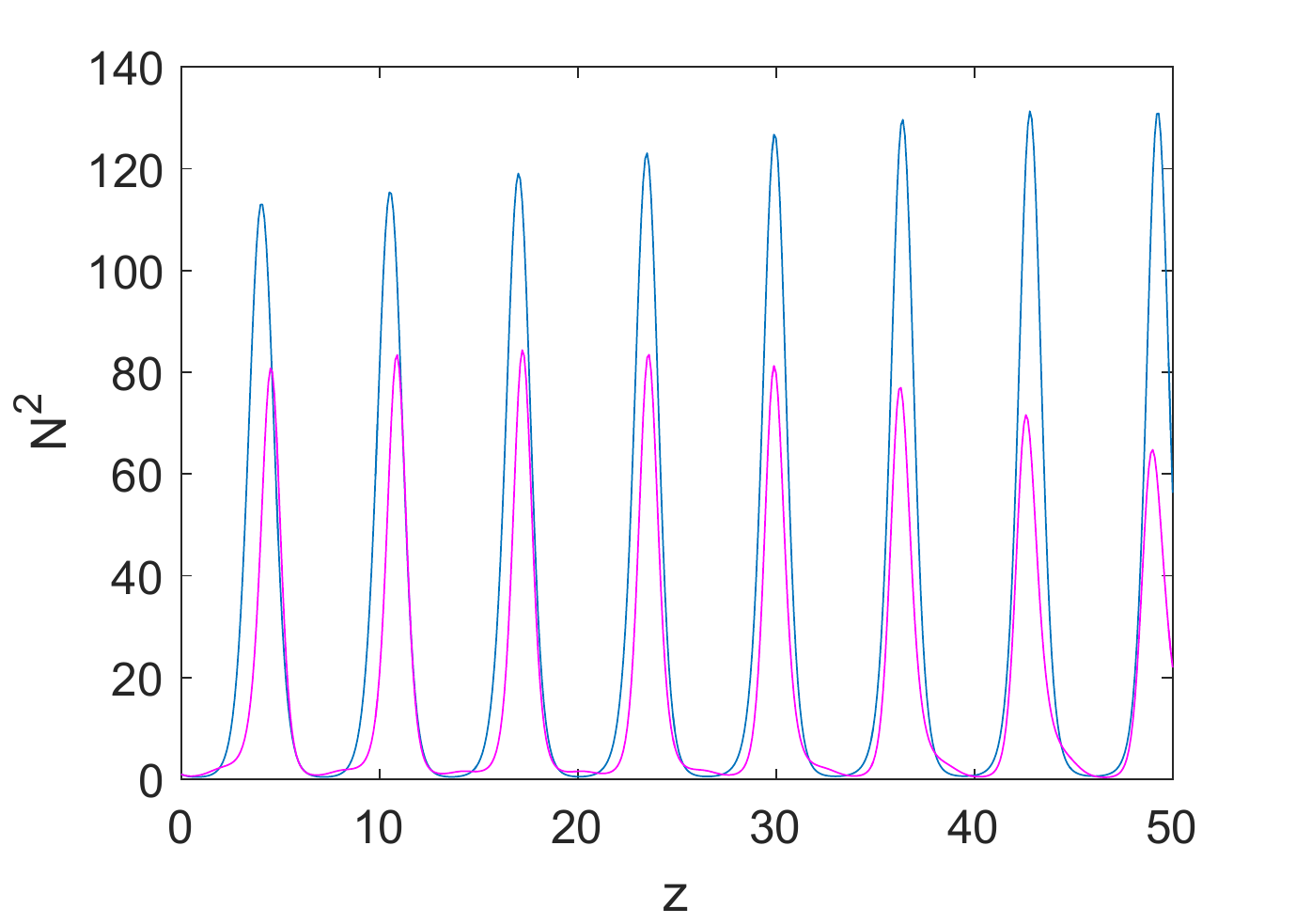}
\caption{Same as figure \ref{fig4}, however, for $q_0=-1$.}
\label{fig5}
\end{figure}

From equations (\ref{eq:Gaussian_approx}) we expect the influence of the gain-loss profile on the beam propagation to be most pronounced where its first derivative with respect to the transversal coordinate is largest, that is near the origin in our example. This can already be observed in figures \ref{fig2} and \ref{fig3} where the propagation for the smallest values of $q_0$ shows the most pronounced modulations. Figure \ref{fig5} shows the same comparisons as Figure \ref{fig4}, however, for a smaller absolute value of $q_0$. We observe that while the differences induced by the gain-loss profile are more pronounced than for the larger initial value of $q_0$ for both $B_0=\rmi$ and $B_0=\frac{\rmi}{2}$, in the second case there are more striking differences developing. These are due to the modulations in the width parameter that are also present in the corresponding Hermitian case, but do not influence the central motion in that case.

\begin{figure}[t]
\centering
\includegraphics[width=0.24\textwidth]{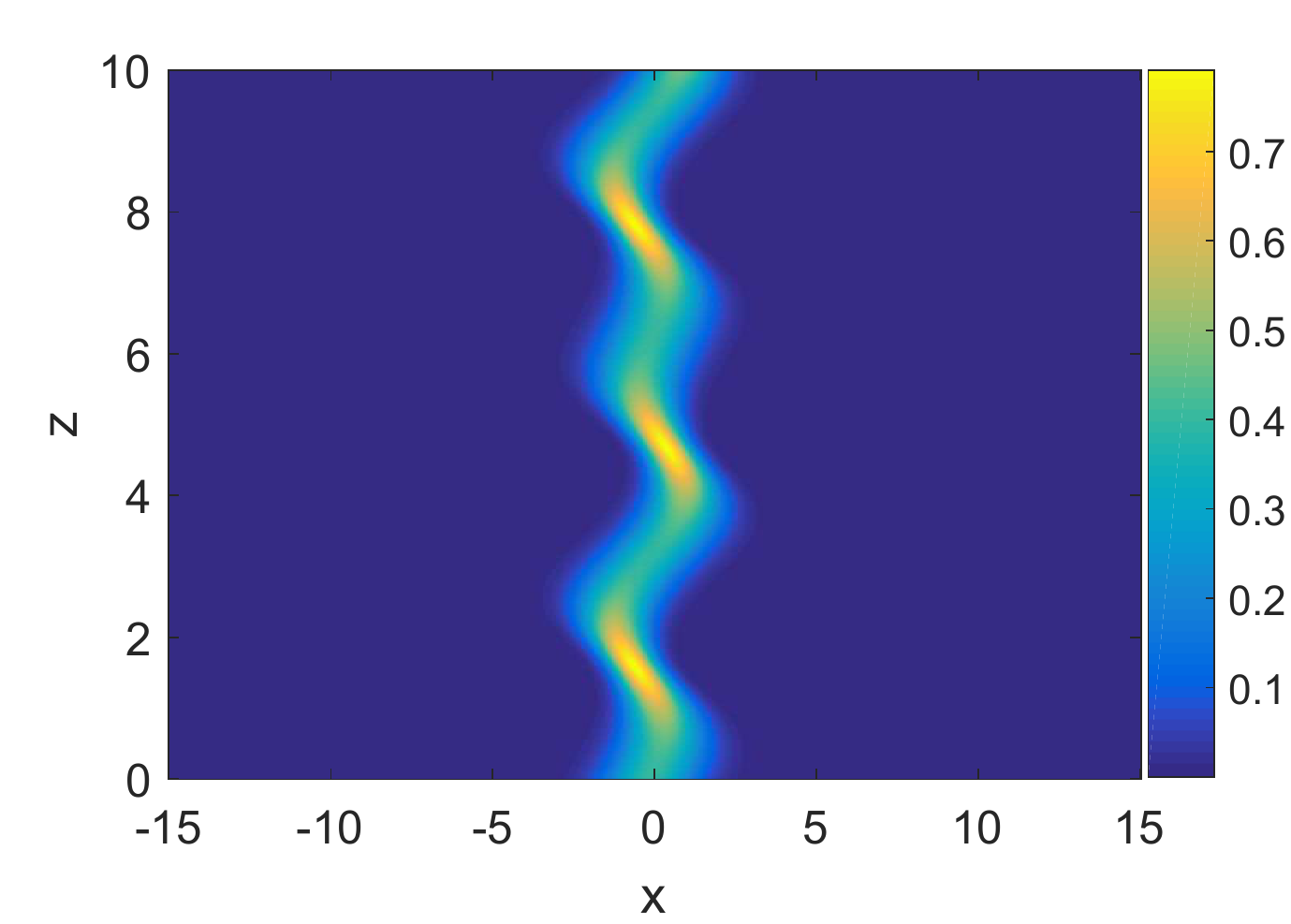}
\includegraphics[width=0.24\textwidth]{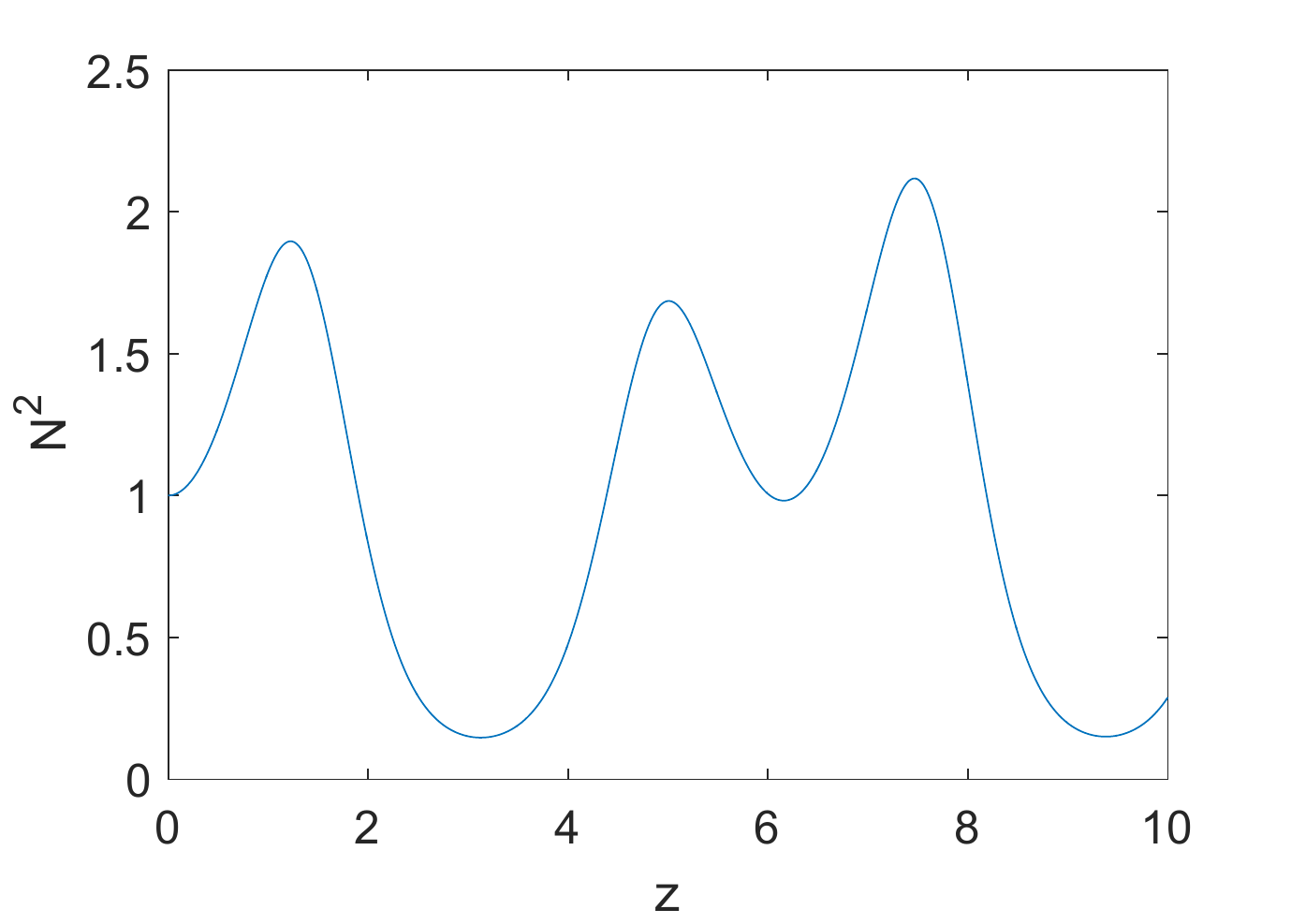}
\includegraphics[width=0.24\textwidth]{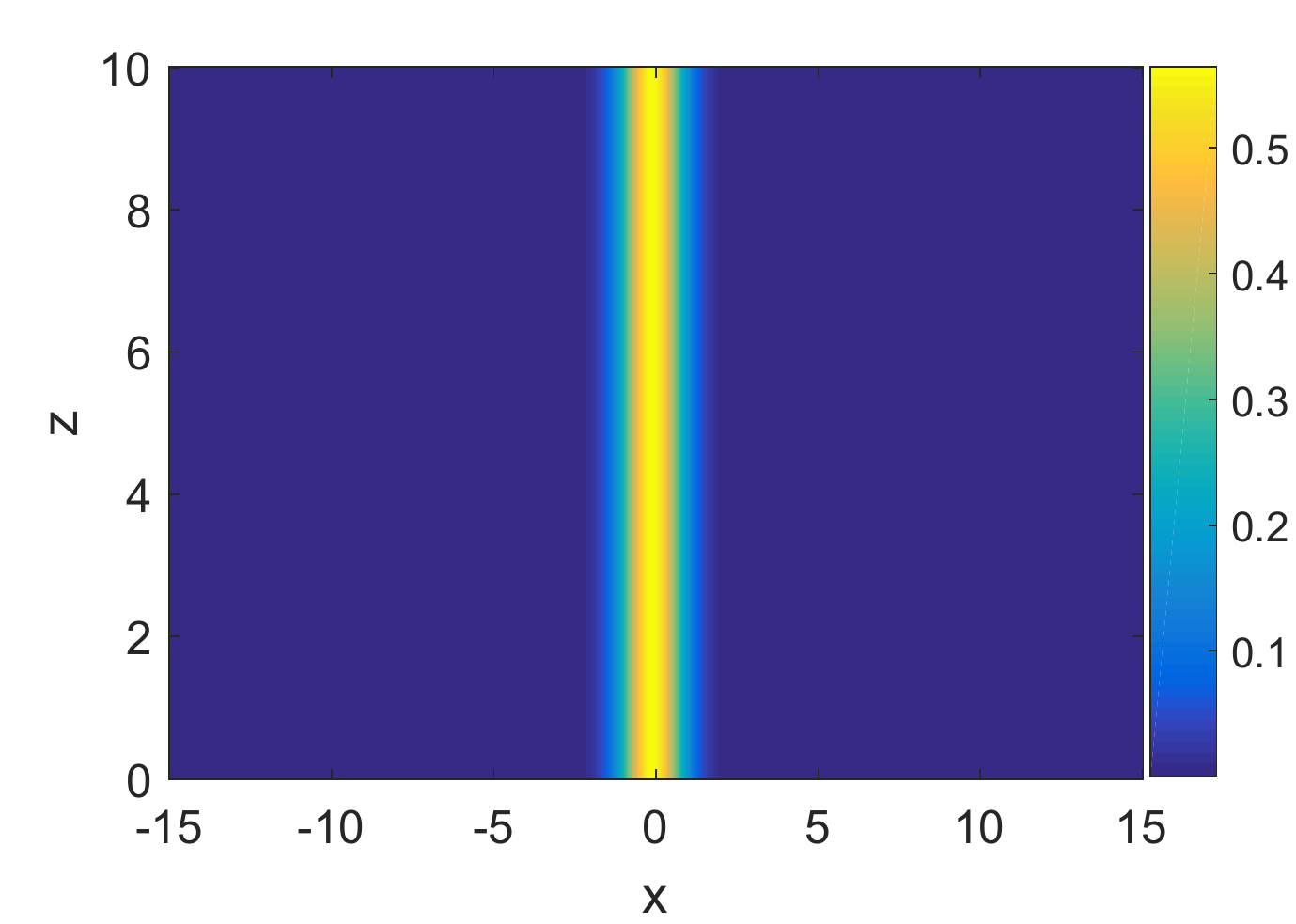}
\includegraphics[width=0.24\textwidth]{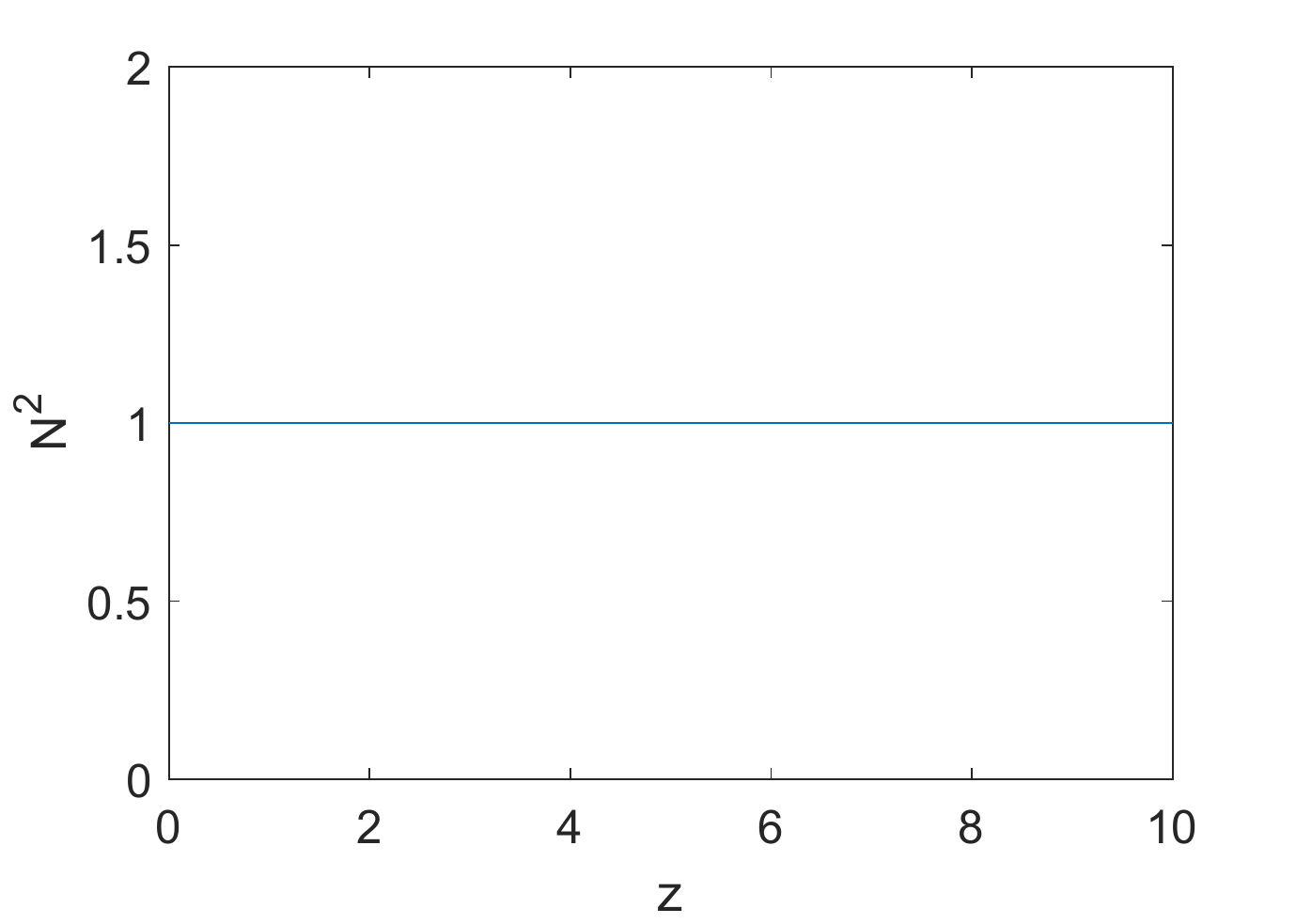}
\includegraphics[width=0.24\textwidth]{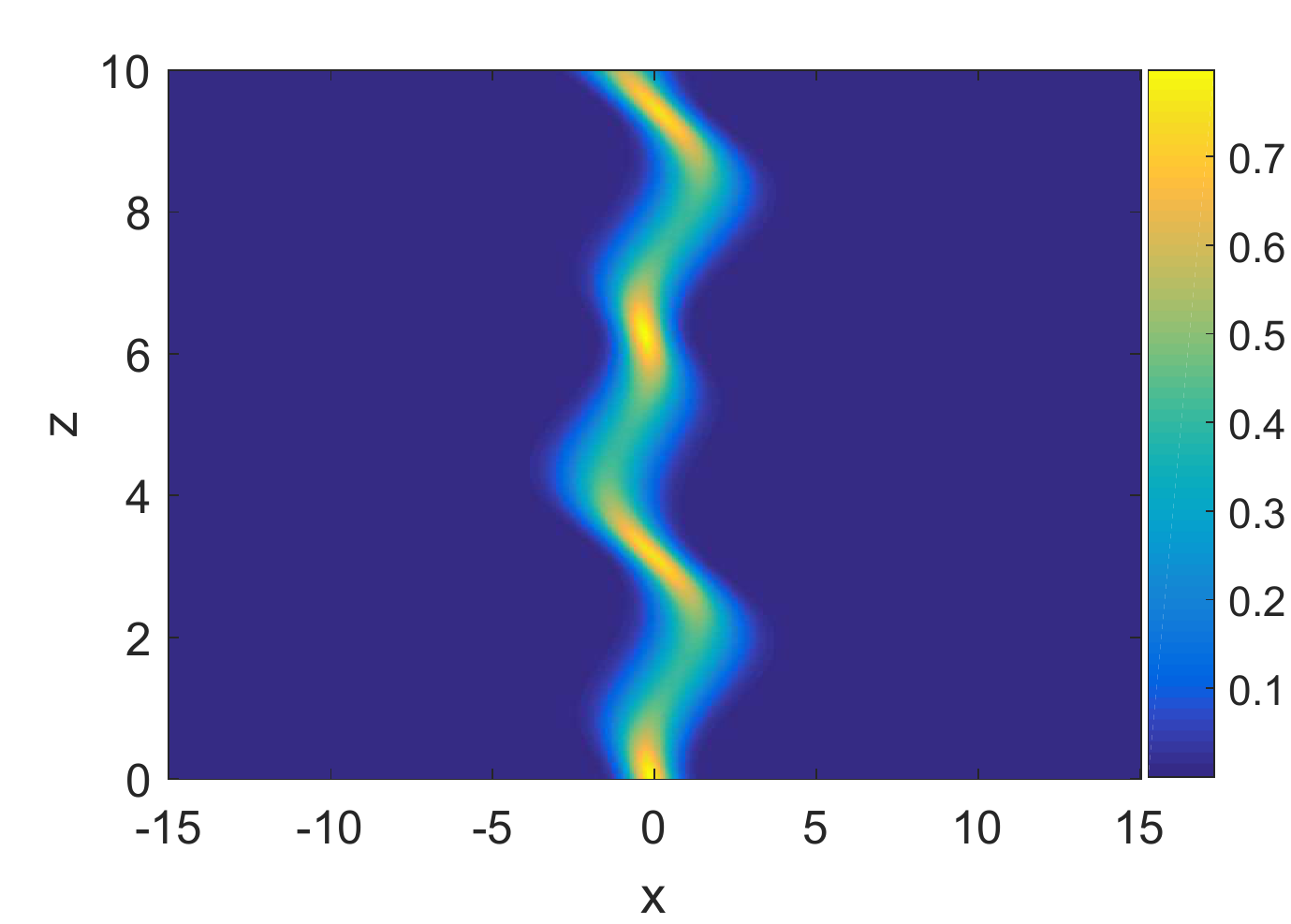}
\includegraphics[width=0.24\textwidth]{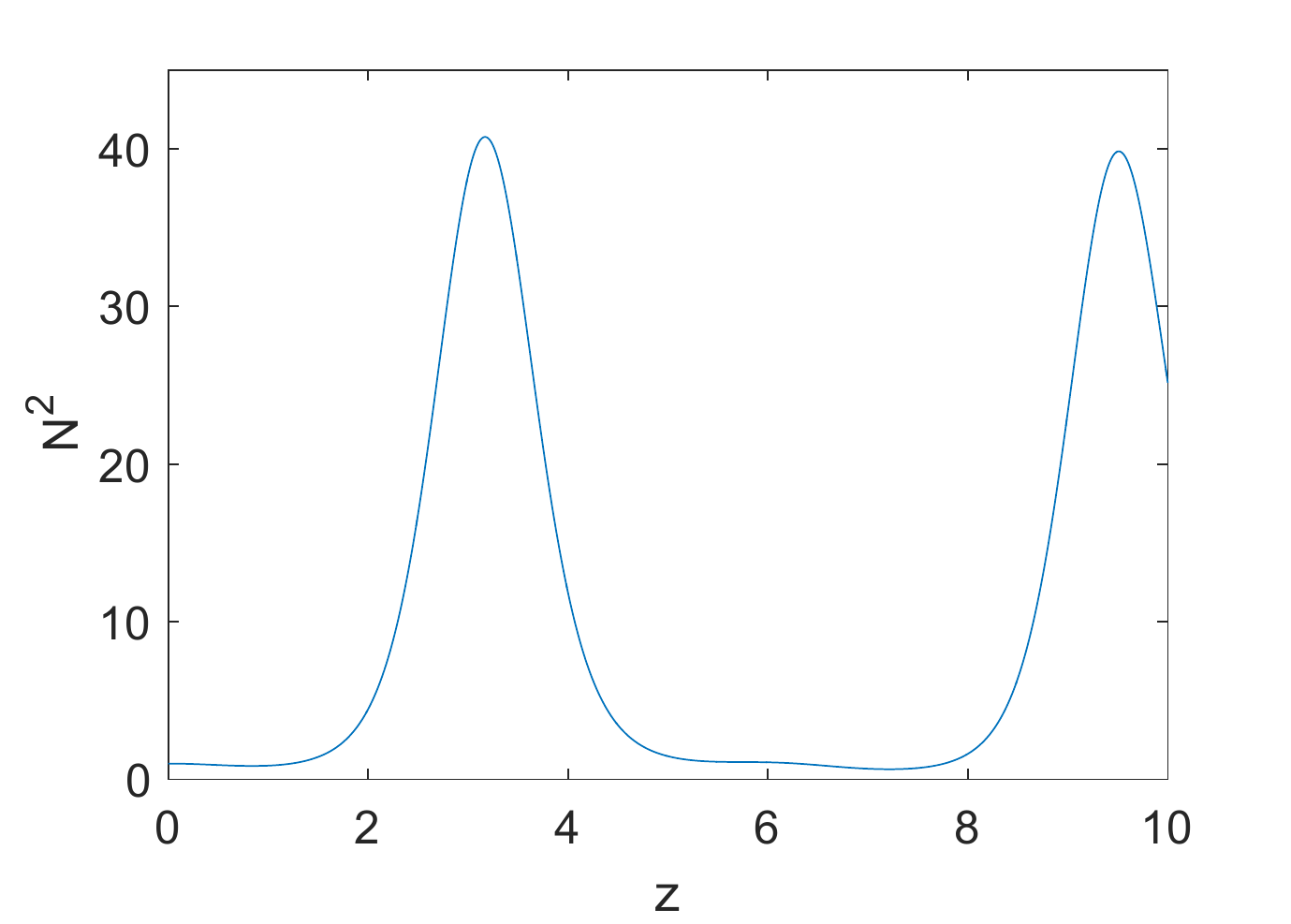}
\includegraphics[width=0.24\textwidth]{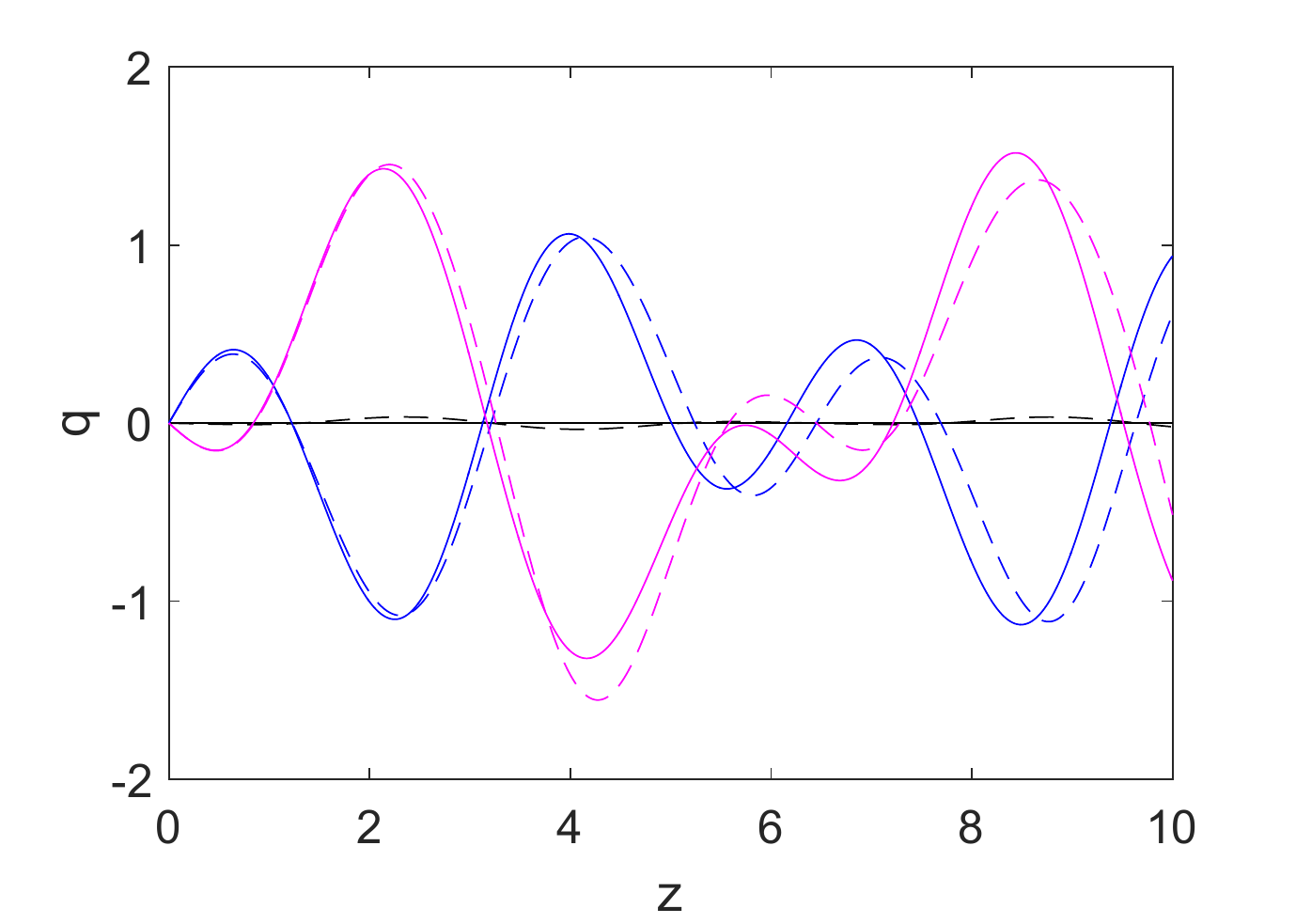}
\caption{(Renormalized) beam propagation in the Gaussian approximation for $\eta=10$, $p_0=-1$, $q_0=0$ and three different values of $B_0$ ($B_0=\frac{\rmi}{2}$ at the top, $B_0=\rmi$ in the middle, and $B_0=2\rmi$ on the bottom.). The left panel depicts the beam propagation, the right panel the norm. At the bottom we show the mean value of the position $q(z)$ (solid lines) in the Gaussian approximation against the exact values computed via the split operator method (dashed lines) for $B_0 = \rmi$ (black), $\frac{\rmi}{2}$ (blue) and $2 \rmi$ (magenta).}
\label{fig6}
\end{figure}

In Figure \ref{fig6} we depict another example where the difference in the initial width leads to drastic differences in the propagation, while all cases are well described by the Gaussian approximation (as can be seen by the direct comparison of the center propagation depicted at the bottom of the figure). In the right panel we depict the propagation of three Gaussian beams with the same initial values of position and momentum, but different initial widths. The right column shows the corresponding norm propagations. The initial momentum is non-zero ($p_0=-1$), nevertheless, for $B=\rmi$, depicted in the middle row, the beam appears stationary. This is due to the fact that the beam is naturally dragged into the gain region, and we have chosen the initial momentum such that it exactly counterbalances this dragging force for $B=\rmi$. Note that it is only possible to create such a stationary solution for this particular value of the width parameter, where the initial Gaussian beam is a very good approximation for the ground state solution of the potential. If the beam is initially wider, as depicted for the example $B_0=\frac{\rmi}{2}$ on the top, the influence of the gain-loss potential is stronger, and the beam starts moving to the right into the gain region and then starts oscillating due to the influence of the real confining potential. If the beam is wider, on the other hand, as shown for the example $B_0=2\rmi$ in the bottom panel of the figure, the influence of the gain-loss profile is effectively decreased. In this case, the initial momentum is strong enough to make the beam move towards the loss region initially, before starting to oscillate.

These effects can  be understood in more detail using the approximative potential
\begin{equation}
\label{eq:pot_exlin}
V = \frac{1}{2}\omega^2 x^2 + \rmi\gamma x,
\end{equation}
which corresponds to the Taylor expansion of the potential \eqref{eq:pot_ex1} around the origin. 
In this approximation the equations of motion for $q, B$ and $N$ simplify to 
\begin{equation}
\begin{aligned}
\ddot{q} = &-\omega^2 q+ \frac{\Re(B)}{\Im(B)}\gamma,\\
\dot B=& - B^{2} - \omega^2,\\
\dot{N} =& \frac{1}{\hbar}\gamma q N,\\
\end{aligned}
\end{equation}
and we can obtain $p$ from $p=\dot{q} - \frac{1}{\Im(B)} \gamma$. The equation for $B$ does no longer depend on $q$, and the equation for $q$ has the form of a forced harmonic oscillator, where the forcing term depends on $B$. Hence, the change of the shape of the wave packet acts like 
a forcing term in the evolution of the centre of the packet. This is true for general systems whenever the wave packet is in a region where the imaginary part of the potential is approximately linear. Once the equation for $q$ is solved, the  norm $N(z)$ is obtained by direct integration 
\begin{equation}
N(z) = N_0\exp\frac{\gamma}{\hbar} \int_0^z q(s)\, \ud s\,\, .
\end{equation}

\begin{figure}[t]
\centering
\includegraphics[width=0.24\textwidth]{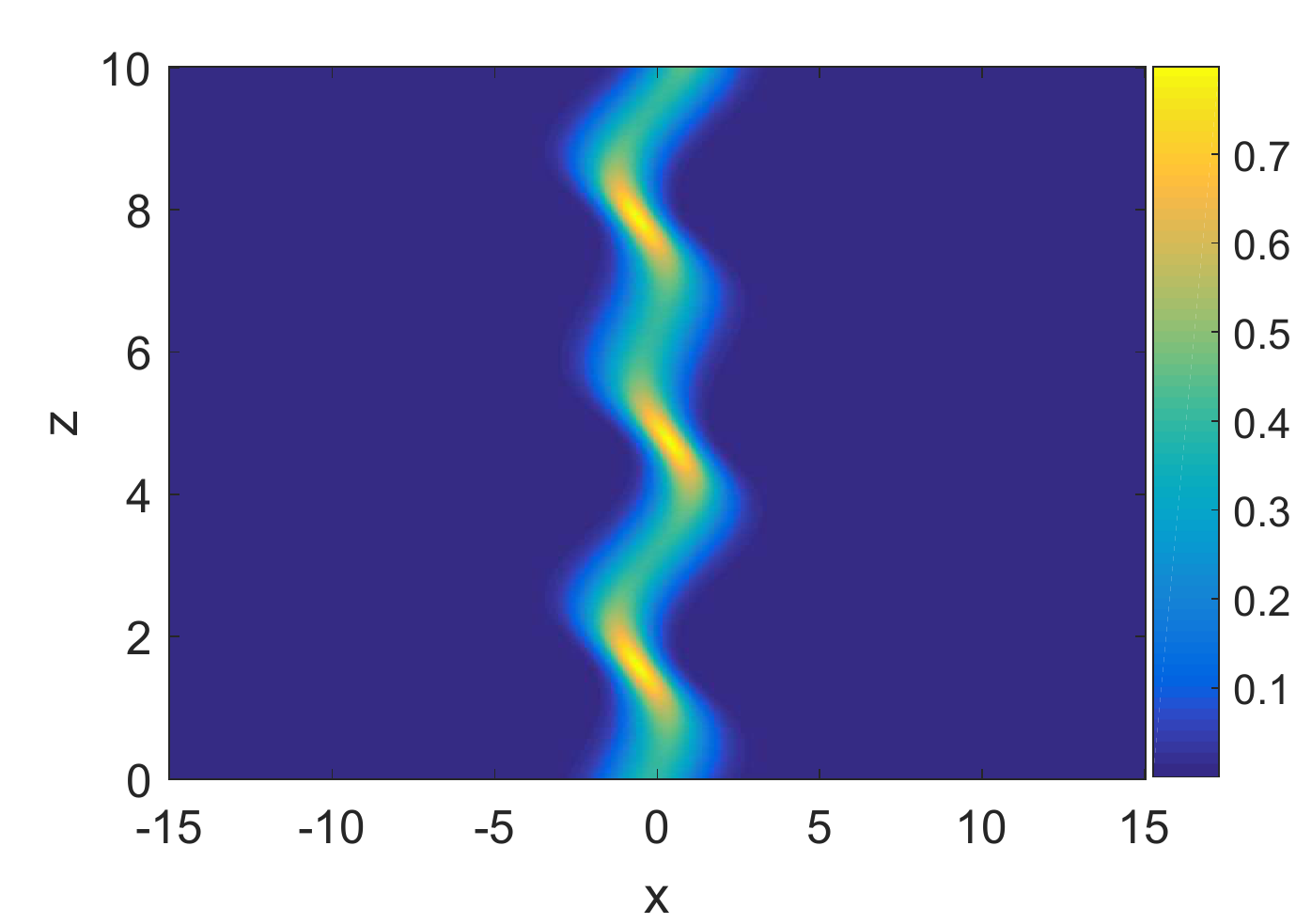}
\includegraphics[width=0.24\textwidth]{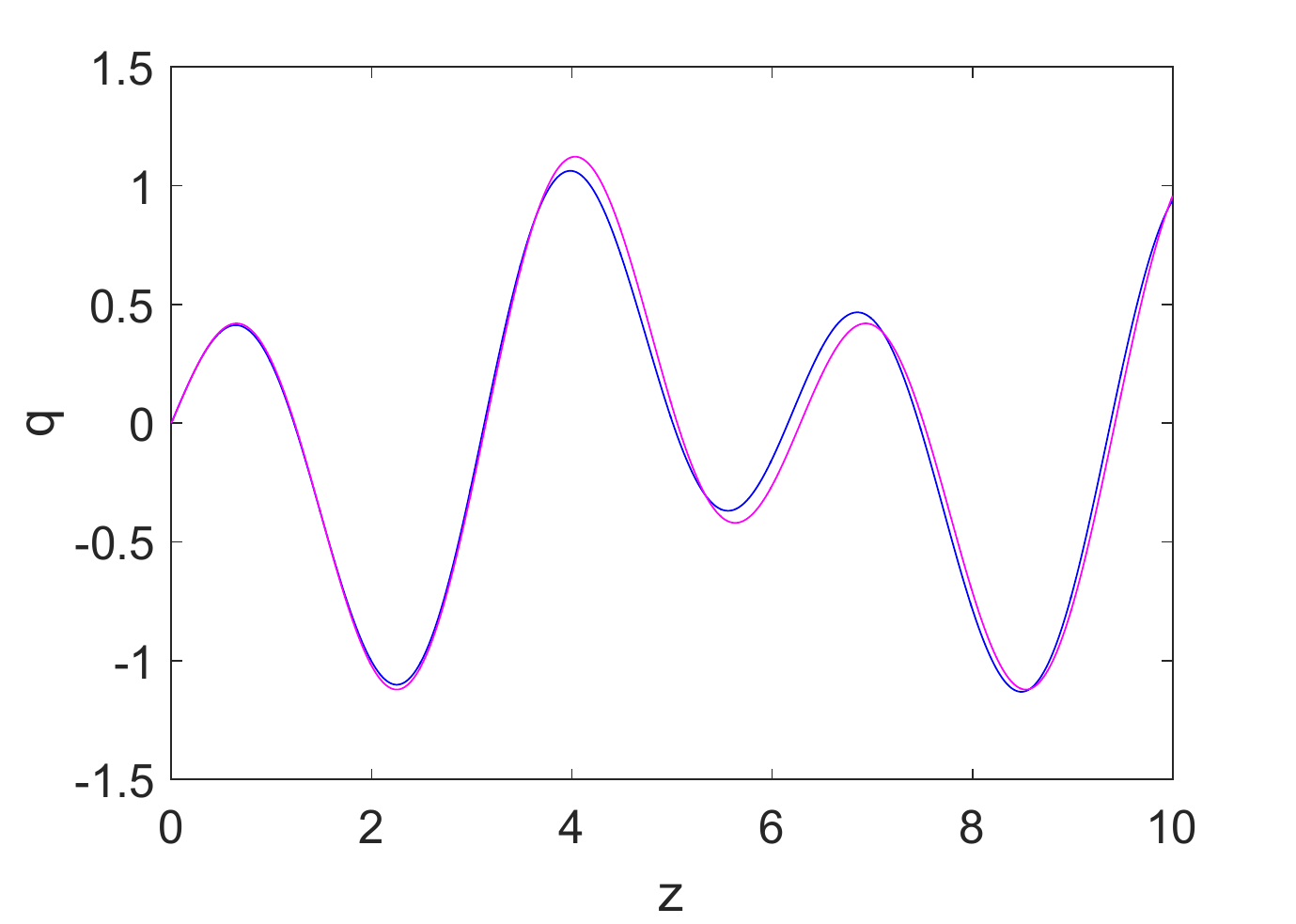}
\includegraphics[width=0.24\textwidth]{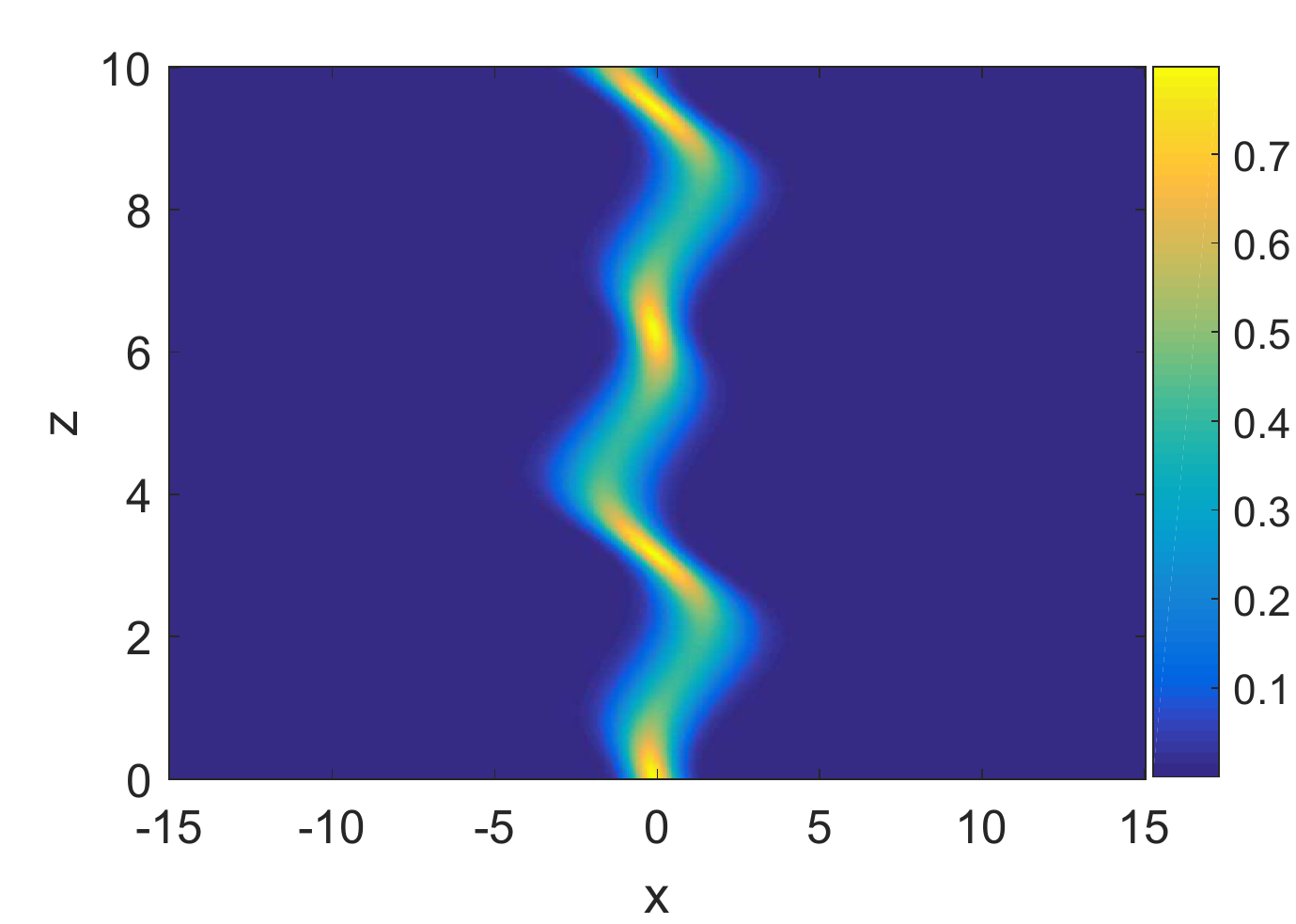}
\includegraphics[width=0.24\textwidth]{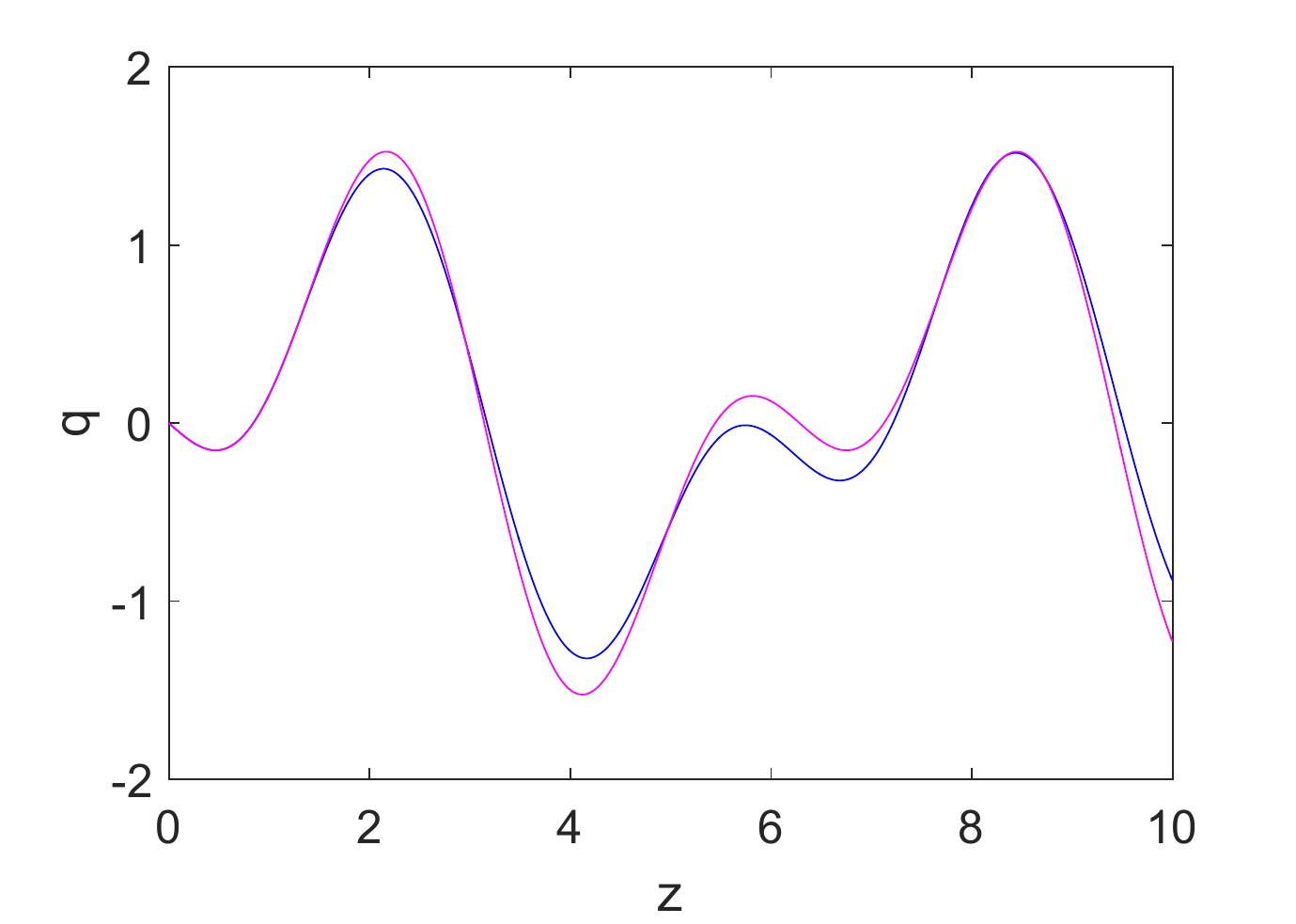}
\includegraphics[width=0.24\textwidth]{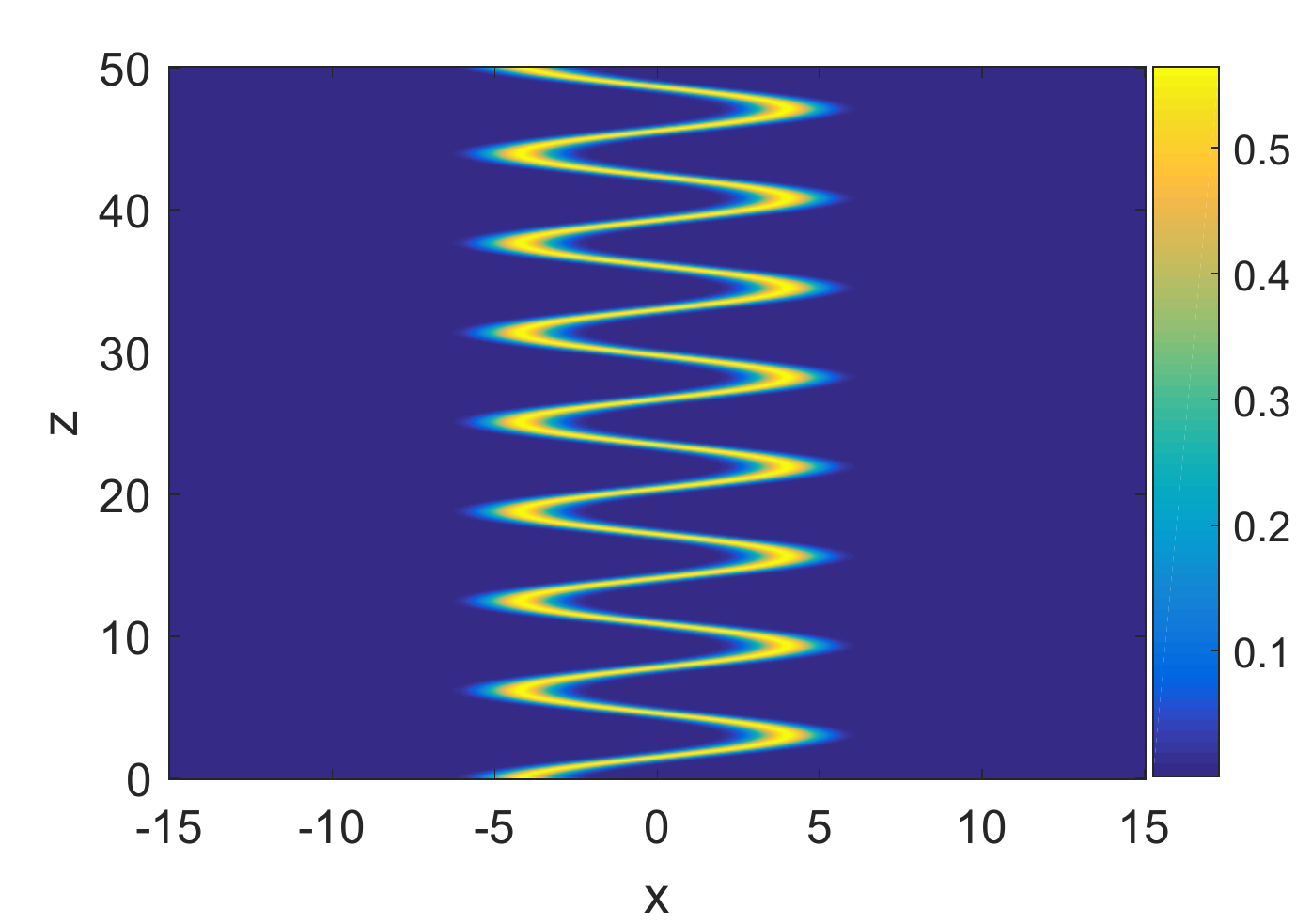}
\includegraphics[width=0.24\textwidth]{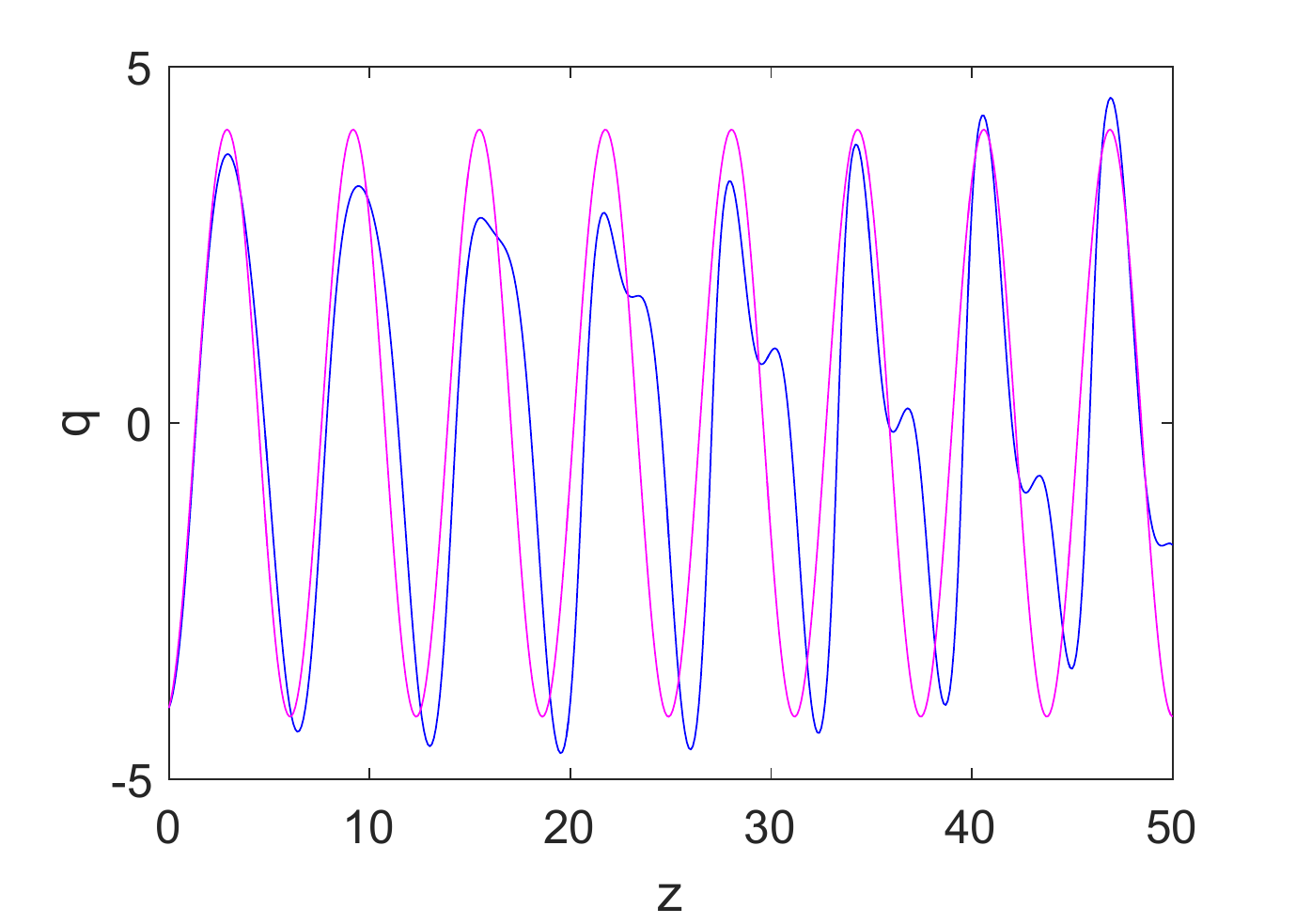}
\caption{(Renormalized) beam propagation in the Gaussian approximation in the potential \eqref{eq:pot_exlin} for parameter values as in Figures \ref{fig6} (top, $B_0 = \frac{\rmi}{2}$, middle: $B_0 = 2\rmi$) and \ref{fig4} (bottom, $B_0 = \rmi$). The left panel shows the renormalized beam propagation, the right panel shows the mean position $q(z)$ of this propagation (magenta line) in comparison to the Gaussian approximation for the exact potential \eqref{eq:pot_ex1} (blue line).}
\label{fig7}
\end{figure}

The equation for $B$ has one stationary solution with $\Im B>0$, namely $B=\rmi \omega$, and for this choice of $B$ the forcing term in the equation for $q$ vanishes and we 
arrive at  a harmonic oscillation with a shift by $\gamma$ in the momentum
\begin{equation}
\begin{aligned}
	q(z) &= q_0 \cos (\omega z) + \frac{p_0 + \gamma}{\omega} \sin (\omega z)  \\
	p(z) &=-\omega q_0\sin(\omega z)+(p_0+\gamma)\cos(\omega z)-\gamma\\
	N(z) &= N_0\rme^{\frac{\gamma}{\hbar} \left(\frac{q_0}{\omega}\sin(\omega z)-\frac{p_0+\gamma}{\omega^2}\cos(\omega z)\right) }.
\end{aligned}
\end{equation}
This is exactly the propagation we have observed in the top panel in Figure \ref{fig5}, as well as the stationary solution in the middle panel of Figure \ref{fig6}.
For a general initial condition $B_0$ with $\Im B_0>0$ the solution to the equation for $B$ is given by 
\begin{equation}
B(z)=\omega \frac{B_0\cos (\omega z) -\omega \sin(\omega z)}{B_0\sin(\omega z)+\omega \cos(\omega z)},
\end{equation}
from which we obtain 
\begin{equation}\label{eq:fracB}
\frac{ \Re(B(z))}{\Im(B(z))}=\frac{|B_0|^2-\omega^2}{2\omega\Im B_0 } \sin(2\omega z)+\frac{\Re B_0}{\Im B_0} \cos(2\omega z),
\end{equation}
that is, the forcing in the $q$ equation has  frequency $2\omega$ and is small if $\Re B_0$ is small and $\Im B_0$ is close to $\omega$. 

The evolution of the center of the wave packet is thus given by 
\begin{equation}
q(z)= a \cos (\omega z) + b\sin (\omega z) -  \frac{\gamma}{\omega^2} \frac{\mathrm{Re}(B(z))}{\mathrm{Im}(B(z))} 
\end{equation}
with $a=\left(q_0 + \frac{\gamma\Re B_0}{\omega^2\Im B_0 }\right)$ and $b=\left(\frac{p_0}{\omega} + \frac{\gamma}{\omega^3} \frac{|B_0|^2}{\Im B_0} \right)$ and 
$\Re B/\Im B$ given by \eqref{eq:fracB}. The norm and the momentum can then be directly obtained as above.  

This explains the modulations with twice the frequency that can be seen on the top and bottom of Figure \ref{fig2}, and in Figures \ref{fig5} and \ref{fig6}, for $B_0\neq\rmi$. The approximation \eqref{eq:pot_exlin} does not only provide a qualitatively but also quantitatively good description of the propagation, as is demonstrated in Figure \ref{fig7}, which depicts the propagations of the two nontrivial cases in Figure \ref{fig6} in the top two rows, using the approximative potential \eqref{eq:pot_exlin}. For a better comparison we also plot the mean values of the position in the right column of the figure. The bottom row depicts the same comparison for a propagation with a larger initial displacement from the origin, as in 
the top row of Figure \ref{fig4}. As expected the details of the propagation are not recovered in this case. 

Let us finish with pointing out a possible application of the dependence of the propagation on the width of the wavepacket as a filtering device. For short propagation distances through a region where the potential is well described by the approximation \eqref{eq:pot_exlin} around the center of the initial wavepacket, 
the position and momentum change according to 
\begin{equation}
\begin{aligned}
	p(z) &=p_0- \left(\omega^2q_0-\gamma \frac{\Re(B_0)}{\Im(B_0)}\right)z\\
	q(z) &= q_0 + \left(p_0+\gamma\frac{1}{\Im(B_0)}\right)z.
\end{aligned}
\end{equation}
That is, for the same initial position and momentum there is an extra shift in the position $q$ that is linear in $\gamma$, i.e., the  slope of the gain-loss profile at $q_0$, and quadratic in the initial width: $q(z)= q_0 + p_0z+2\gamma(\Delta q)^2 z$. This effect could be used to spatially separate Gaussian beams with different widths. If the width parameter of the beam also has a non-vanishing real part, this leads to an additional shift in the momentum, which translates into the angle in optical applications. At the same time the overall norm is only changed linearly according to 
\begin{equation}
	N(z) =N_0 (1+\gamma q_0 z),
\end{equation}
for short propagation distances.

\section{Summary and Conclusion}
We have derived semiclassical equations of motion for Gaussian beams propagating in waveguides in the paraxial approximation in the presence of gain and loss. In the absence of losses this leads to Hamiltonian motion of the center with a time dependent width. In the presence of gain and/or loss, however, the width of the beam influences the central motion. We have demonstrated that this Gaussian approximation can capture typical features of beam propagation in gain-loss wave guides such as power oscillations, and can accurately describe the propagation if the refractive index is well described by its Taylor expansion up to second order on length scales given by the typical widths of the Gaussian beam. Finally we have demonstrated how the dependence on the width could be used as a filtering device. 

\section*{Acknowledgment}
E.M.G. acknowledges support from the Royal Society via a University Research Fellowship (Grant. No. UF130339), and a L'Or\'eal-UNESCO for Women in Science UK and Ireland Fellowship. A.R. acknowledges support from the Engineering and Physical Sciences Research Council via the Doctoral Research Allocation Grant No. EP/K502856/1.


\begin{thebibliography}{10}
\bibitem{HeppHeller}
K.~Hepp,  Com. Math. Phys.  {\bf 35}  (1974)   265; E.~J. Heller,  J.
  Chem. Phys.  {\bf 62}  (1975)   1544

\bibitem{Litt86}
R.~G. Littlejohn,  Phys.
  Rep.  {\bf 138}  (1986)   193

\bibitem{Hell_dyn}
E.~J. Heller,  J. Chem. Phys.  {\bf 75}  (1981)   2923; D.~Huber and E.~J. Heller,    J. Chem. Phys.  {\bf 89}  (1988)   4752; D.~Huber, S.~Ling, D.G. Imre, and E.~J. Heller,  J. Chem. Phys.  {\bf 90}  (1989)   7317

\bibitem{Long09b}
S.~Longhi,  Laser \& Photon. Rev.  {\bf 3}  (2009)   243

\bibitem{PT}
C.~M. Bender, S.~Boettcher, and P.~N. Meisinger,  J. Math. Phys.  {\bf 40}  (1999)   2201
  
\bibitem{PT_optics} R.~El-Ganainy, K.~G. Makris, D.~N. Christodoulides, and Z.~H. Musslimani,  Opt.  Lett.  {\bf 32}  (2007)   2632; K.~G.~Makris, R.~El-Ganainy, D.~N.~Christodoulides, and Z.~H.~Musslimani, Phys. Rev. Lett. \textbf{100}, 103904 (2008); S.~Klaiman, U.~G\"unther, and N.~Moiseyev,  
Phys. Rev. Lett. \textbf{101}, 080402 (2008); A.~Guo, G.~J.~Salamo, D.~Duchesne, R.~Morandotti, M.~Volatier-Ravat, 
V.~Aimez, G.~A.~Siviloglou, and D.~N.~Christodoulides, 
Phys. Rev. Lett. \textbf{103}, 093902 (2009); S. Longhi, Phys. Rev. Lett {\bf 103}, 123601 (2009); 
C.~E.~R\"uter, K.~G.~Makris, R.~El-Ganainy, D,~N,~Christodoulides,
and D.~Kip, 
Nat. Phys. \textbf{6}, 192 (2010); 
E.~M. Graefe and H.~F. Jones, Phys. Rev. A  {\bf 84}
   (2011)   013818; A.~Regensburger, C.~Bersch, M-A Miri, G.~Onishchukov, D.~N. Christodoulides,  and U.~Perschel,  Nature  {\bf 488}  (2012)   167
  
  \bibitem{11nhcs}
E.~M. Graefe and R.~Schubert,  Phys. Rev. A  {\bf 83}  (2011)   060101(R); E.~M. Graefe and R.~Schubert,  J. Phys. A  {\bf 45}  (2012)
  244033

\end{thebibliography}
\end{document}